\documentclass[useAMS, usenatbib]{mnras}
\usepackage{color}
\usepackage{booktabs}   % Fancier tables
\usepackage{tabularx}
\usepackage{float}
\usepackage{totcount}
\usepackage{cuted}

\usepackage[dvipsnames, table]{xcolor} % adds colors to colorspace (https://en.wikibooks.org/wiki/LaTeX/Colors)
\usepackage{graphicx}
\usepackage{multirow}
\usepackage[section]{placeins}   % Force figures to stay in their sections (tex.stackexchange.com/q/279/22806)
\usepackage{subfig}
\usepackage{widetext}
\usepackage{upgreek}    % allow alternative greek letters like $\uptau$
\usepackage{enumitem}   % customize enumerate symbols
\setlist[enumerate]{leftmargin=*}   % Set left-margin of enumerate lists to match the edge
\setlist[itemize]{leftmargin=*}   % Set left-margin of enumerate lists to match the edge

\usepackage{lipsum} % generate dummy text
\usepackage{amsmath}
\usepackage{amssymb}
\usepackage{subfiles}    % be able to compile the included subfiles

\usepackage{txfonts}     %more compact fonts closer to the ones used by MNRAS

\graphicspath{ {figs/} }

\newcommand{\tr}[1]{\textrm{#1}}
\newcommand{\trt}[1]{\textrm{\tiny{#1}}}

\newcommand{\msol}{\tr{M}_{\odot}}

\newcommand{\lr}[1]{\left(#1\right)}
\newcommand{\E}[1]{\times\nobreak10^{#1}}

\newcommand{\secref}[1]{\textsection\ref{#1}}
\newcommand{\figref}[1]{Fig.~\ref{#1}}
\newcommand{\refeq}[1]{{Eq.~\ref{#1}}}      % no parentheses around equation numbers
\newcommand{\tabref}[1]{{Table~\ref{#1}}}

\newcommand{\rinfl}{\mathcal{R}_\trt{infl}}

\newcommand{\rh}{R_\trt{h}}    % Hard radius
\newcommand{\ayr}{A_{\trt{yr}^{-1}}}        % GWB amplitude normalization at 1/yr
\newcommand{\pyr}{\textrm{yr}^{-1}}
\newcommand{\pc}{\mathrm{pc}}

\newcommand{\SNR}{\tr{S}/\tr{N}}   % Signal-to-Noise Ratio SNR

\newcommand{\comdist}{d_\trt{c}}   % 'Comoving Distance'
\newcommand{\lumdist}{d_\trt{L}}   % 'Comoving Distance'
\newcommand{\astropy}{\texttt{Astropy}}
\newcommand{\matplotlib}{\texttt{matplotlib}}
\newcommand{\numpy}{\texttt{NumPy}}
\newcommand{\scipy}{\texttt{SciPy}}
\newcommand{\ipython}{\texttt{ipython}}

\newcommand{\blank}{\, ... \,}

\def\aj{Astron. J.}              		   		% Astronomical Journal
\def\apj{Astrophys. J.}       		        	 	% Astrophysical Journal
\def\apjl{Astrophys. J. Lett.}             		% Astrophysical Journal, Letters

\def\pasa{PASA}
\def\apjs{Astrophys. J., Suppl. Ser.}            % Astrophysical Journal, Supplement
\def\mnras{Mon. Not. R. Astron. Soc.}        % Monthly Notices of the RAS
\def\prd{Phys. Rev. D}      				% Physical Review D
\def\nat{Nature}              				% Nature
\def\aap{Astron. Astrophys.}               		% Astronomy and Astrophysics

\newcommand{\heavy}{\textit{heavy}}  % 'Heavy' Binaries
\newcommand{\major}{\textit{major}}  % 'Major-Mergers' Binaries
\newcommand{\heavymajor}{\heavy~\&~\major}
\newcommand{\heavydef}{{\textit{heavy} ($M > 10^{9} \, \msol$)}}  % 'Heavy' Binaries
\newcommand{\majordef}{{\textit{major} ($\mu > 0.1$)}}  % 'Major-Mergers'

\newcommand{\frest}{f_\trt{r}}
\newcommand{\fobs}{f}
\newcommand{\fharm}{f_\trt{h}}

\newcommand{\hc}{h_\trt{c}}
\newcommand{\hs}{h_\trt{s}}
\newcommand{\hsc}{h_\trt{s,circ}}

\newcommand{\volfactor}{\Lambda_{ij}}
\newcommand{\poisson}{\mathcal{P}}

\newcommand{\iprime}{IPTA$'$}
\newcommand{\iptazero}{IPTA$_0$}
\newcommand{\iprimeone}{IPTA$'_1$}
\newcommand{\iprimerap}{IPTA$'_\mathrm{rap}$}

\newcommand{\frefill}{\mathcal{F}_\trt{refill}}

\newcommand{\mchirp}{\mathcal{M}}     % Chirp-mass 
\newcommand{\volill}{V_\trt{ill}}   % Illustris volume

\newcommand{\bigt}{\scalebox{1.2}{\ensuremath{\uptau}}}
\newcommand{\tfhard}{\bigt_\tr{h}^{f}}
\newcommand{\thard}{\bigt_\tr{h}}
\newcommand{\thubble}{\bigt_\tr{Hubble}}
\newcommand{\thardgw}{\bigt_\tr{gw}}

\newcommand{\lgw}{L_\tr{GW}}
\newcommand{\lgwc}{L_\tr{GW,circ}}

\newcommand{\egw}{\varepsilon_\trt{GW}}   % energy in GW

\newcommand{\closesim}{{\sim}}

\newcommand{\aul}{A_\trt{yr$^{-1}$,ul}}

\newcommand{\scalenoise}{\lambda_\tr{noise}}
\newcommand{\simclose}{{\sim}}

\defcitealias{hkm09}{HKM09}
\defcitealias{bbr80}{BBR80}
\defcitealias{paper1}{KBH-16}

\def\biblio{\bibliographystyle{mnras}\bibliography{\main/refs}}  

\newtotcounter{citnum} %From the package documentation
\def\oldbibitem{} \let\oldbibitem=\bibitem
\def\bibitem{\stepcounter{citnum}\oldbibitem}

\title[GWB Predictions from Illustris MBHB]{The Gravitational Wave Background from Massive Black Hole Binaries in Illustris: spectral features and time to detection with pulsar timing arrays}
\author[L.Z.~Kelley et al.]{Luke Zoltan Kelley$^{1}$\thanks{E-mail:lkelley@cfa.harvard.edu},
	Laura Blecha$^{2}$,
	Lars Hernquist$^{1}$,
	Alberto Sesana$^{3}$,
	Stephen R. Taylor$^{4}$
\\
$^{1}$ Harvard University, Center for Astrophysics \\
$^{2}$ University of Maryland \\
$^{3}$ University of Birmingham \\
$^{4}$ Jet Propulsion Laboratory, California Institute of Technology
}

\begin{document}

% Override 'biblio' command for this document (suppress subfiles bibliographies)
\def\biblio{}

%\date{Accepted 1988 December 15. Received 1988 December 14; in original form 1988 October 11}
\pagerange{\pageref{firstpage}--\pageref{lastpage}} \pubyear{2014}

\maketitle
\label{firstpage}

\begin{abstract} 
Pulsar Timing Arrays (PTA) around the world are using the incredible consistency of millisecond pulsars to measure low frequency gravitational waves from (super)Massive Black Hole (MBH) binaries.  We use comprehensive MBH merger models based on cosmological hydrodynamic simulations to predict the spectrum of the stochastic Gravitational-Wave Background (GWB).  We use real Time-of-Arrival (TOA) specifications from the European, NANOGrav, Parkes, and International PTA (IPTA) to calculate realistic times to detection of the GWB across a wide range of model parameters.  In addition to exploring the parameter space of environmental hardening processes (in particular: stellar scattering efficiencies), we have expanded our models to include eccentric binary evolution which can have a strong effect on the GWB spectrum.  Our models show that strong stellar scattering and high characteristic eccentricities enhance the GWB strain amplitude near the PTA sensitive ``sweet-spot" (near the frequency $f = 1 \, \pyr$), slightly improving detection prospects in these cases.  While the GWB \textit{amplitude} is degenerate between cosmological and environmental parameters, the location of a spectral turnover at low frequencies ($f \lesssim 0.1 \, \pyr$) is strongly indicative of environmental coupling.  At high frequencies ($f\gtrsim 1 \, \pyr$), the GWB spectral index can be used to infer the number density of sources and possibly their eccentricity distribution.  Even with merger models that use pessimistic environmental and eccentricity parameters, if the current rate of PTA expansion continues, we find that the International PTA is highly likely to make a detection within about 10 years.

\end{abstract}

\begin{keywords}
quasars: supermassive black holes, galaxies: kinematics and dynamics
\end{keywords}

% =================================================================================================
% = = = = = = = = = = = = = = = = = = = =   CORE / BODY   = = = = = = = = = = = = = = = = = = = = =
% =================================================================================================

% Introduction
% ------------

% ==================================================================
% =======================   INTRODUCTION   =========================

\section{Introduction}
\label{sec:intro}

% Pulsar Timing Arrays --
Pulsar Timing Arrays (PTA) are expected to detect Gravitational Waves \citep[GW;][]{sazhin1978, detweiler1979, romani1983} from stable binaries of (super)-Massive Black Holes \citep[MBH;][]{rajagopal1995, wyithe2003, phinney2001}.  These arrays use correlated signals in the consistently timed pulses from millisecond pulsars to search for low-frequency ($\lesssim 10 \, \pyr$) perturbations to flat space-time \citep{hellings1983, foster1990}.  There are currently three independent PTA searching for GW signals: the North-American Nanohertz Observatory for Gravitational waves \citep[NANOGrav;][]{mclaughlin2013}, the European PTA \citep[EPTA;][]{kramer2013}, and the Parkes PTA \citep[PPTA;][]{manchester2013a}. Additionally, the International PTA \citep[IPTA,][]{hobbs2010} is a collaboration which aims to combine the data and expertise from each independent group.

% Upper Limits, and Consistency with Models --
Comparable upper limits on the presence of a stochastic Gravitational Wave Background (GWB) have been calculated by the EPTA \citep{lentati2015}, NANOGrav \citep{arzoumanian2015b}, PPTA \citep{shannon2015} and IPTA \citep{verbiest2016}.  These upper limits are already astrophysically informative in that much of the previously predicted parameter space is now in tension with observations, and there are suggestions that some models are excluded \citep{shannon2015}.  Many previous GWB models have assumed that most or all of the MBH pairs formed after the merger of their host galaxies are able to quickly reach the `hard binary' phase\footnote{`Hard' binaries are distinguished by, and important because, scattering interactions tend to further harden the binary \cite[e.g.][]{hut1983}.} ($\lesssim 10 \, \mathrm{pc}$) and eventually coalesce due to GW emission \citep[e.g.][]{wyithe2003, jaffe2003, sesana2013}.  These models, which assume GW-only driven evolution and produce purely power-law GWB spectra, likely over-predict the GWB energy in that: 1) perhaps a substantial fraction of MBH Binaries stall at galactic scales ($\sim$ kpc), or before reaching the small separations ($\closesim 10^{-3}$--$10^{-1} \, \mathrm{pc}$) corresponding to the PTA sensitive band \citep[e.g.][]{mcwilliams2014}; and 2) significant `attenuation' of the GW signal may exist due to environmental processes (non-GW hardening, due to stellar scattering or coupling with a circumbinary gaseous disk) which decrease the amount of time binaries spend in a given frequency interval \citep[e.g.][]{kocsis2011, sesana2013, ravi2014, rasskazov2016}.

% New Merger Models, and Time to Detection --
Some recent GW-only models predict lower signal levels because of differing cosmological assumptions (i.e.~galaxy-galaxy merger rates, the mass functions of MBH, etc) which produce different distributions of MBH binaries \citep[e.g.][]{roebber2016, sesana2016}, eliminating the tension with PTA upper limits.  More comprehensive models have also been assembled which take into account binary-stalling and GW-attenuation \citep[e.g.][]{ravi2014}.  Some of these models suggest that the GWB is only just below current observational sensitivities, which begs the question, `\textit{how long until we make a detection?}'  Recently, PTA detection statistics conveniently formalized in \citet{rosado2015} have been used by \citet{taylor2016} to calculate times to detections for purely power-law, GW-only GWB models with a full range of plausible GWB amplitudes.  \citet{vigeland2016} also calculate detection statistics using a more extensive suite of broken power-laws to model the effects of varying environmental influences.

% Our Models --
In \citet[][hereafter `\citetalias{paper1}']{paper1} we construct the most comprehensive MBHB merger
models to date, using the self-consistently derived population of galaxies and MBH from the Illustris
cosmological, hydrodynamic simulations \citep[\secref{sec:meth_ill}, e.g.][]{vogelsberger2014a, genel2014}.
The MBHB population is post-processed using semi-analytic models of GW emission in addition to
environmental hardening mechanisms that are generally \textit{required} for MBHB to
reach small separations within a Hubble time \citep[e.g.][]{bbr80, milosavljevic2003}, and emit GW in
PTA-sensitive frequency bands.

% Paper Outline --
In this paper, we introduce the addition of eccentric binary evolution to our models, and explore
its effects on the GWB.  To produce more realistic GWB spectra, we complement our previous semi-analytic (SA)
calculations with a more realistic, Monte-Carlo (MC) technique. 
Using our merger models and resulting spectra, we calculate realistic times to detection for each
PTA following \citet{rosado2015} \& \citet{taylor2016}.  In \secref{sec:meth} we describe our MBHB, GWB
and PTA models.  Then in \secref{sec:results} we describe
the effects of eccentricity on binary evolution (\secref{sec:res_ecc}) and the GWB spectrum
(\secref{sec:res_gwb}) including comparisons between the SA and MC calculations, and finally our
predictions for times to GWB detections (\secref{sec:res_pta}).

% Create a bibliography here, only if just this file is being compiled/built.
\biblio{}

% Illustris
% ---------

% =============================================================
% =======================   METHODS   =========================

\section{Methods}
\label{sec:meth}
Our simulations use the coevolved galaxies and MBH particles from the Illustris cosmological,
hydrodynamic simulations \citep{vogelsberger2013, torrey2014, vogelsberger2014b, genel2014, sijacki2015}
run using the Arepo `moving-mesh' code \citep{springel2010}.
Our general procedure of extracting MBH, their merger events and their galactic environments are
described in detail in \citetalias{paper1}.  Here, we give a brief overview of our methods
(\secref{sec:meth_ill}) and the improvements made to include eccentric binary evolution
(\secref{sec:meth_eccen}).  We then describe the methods by which we calculate GW signatures
(\secref{sec:meth_gw}) and realistic detection statistics for simulated PTA (\secref{sec:meth_pta}).

% Illustris MBH Mergers
% ---------------------
\subsection{Illustris MBH Mergers and Environments}
\label{sec:meth_ill}
The Illustris simulation is a cosmological box of $106.5 \, \mathrm{Mpc}$ on a side (at $z=0.0$)
containing moving-mesh gas cells, and particles representing stars, dark matter, and MBH.
All of the Illustris data is publicly available online \citep{nelson2015}.  MBH
are `seeded' with a mass of $1.42\E{5} \, \msol$ into halos with masses above $7.1\E{10} \, \msol$
\citep{sijacki2015}, where they accrete gas from the local environment and grow over time.
As they develop, they proportionally deposit energy back into the local environment
\citep{vogelsberger2013}.
When two MBH particles come within a gravitational smoothing length of
one another (typically on the order of a $\mathrm{kpc}$) a `merger' event is recorded.  From those
mergers we identify the constituent MBH and the host galaxy in which they subsequently reside.
From the host galaxy,
density profiles are constructed which are used to determine the environment's influence on the
MBHB merger process.  The simulations used here, as in \citetalias{paper1}, are semi-analytic models
which integrate each binary (independently) from large-scale separations down to eventual coalescence
based on prescriptions for GW- and environmentally- driven hardening.

% Eccentric Binary Evolution
% --------------------------
\subsection{Models for Eccentric Binary Evolution}
\label{sec:meth_eccen}
We implement four distinct mechanisms which dissipate orbital energy
and `harden' the MBHB (as in \citetalias{paper1}):

	\begin{itemize}
	% Dynamical Friction -----
    \item
    \textbf{Dynamical Friction (DF,} dominant on $\simclose \mathrm{kpc}$ scales) is implemented following \citet{chan43} and \citet{bt87}, based on the local density (gas and dark matter) and velocity dispersion.  The mass of the decelerating object is taken as the mass of the secondary MBH along with its host galaxy.  We assume a model for tidal stripping such that the decelerating mass decreases as a power-law from the combined mass, to that of only the secondary MBH, over the course of a dynamical time\footnote{The dynamical time used is that of the primary's host galaxy.  This corresponds to the `Enh-Stellar' model from \citetalias{paper1}.}.  With the addition of eccentric binary evolution, we make the approximation that DF does not noticeably affect the eccentricity distribution of binaries \citep[e.g.][]{colpi1999, vandenBosch1999, hashimoto2003}, and that the semi-major axis remains the relevant distance scale.  Equivalently, the ``initial'' eccentricities in our models can be viewed as the eccentricity once binaries enter the stellar scattering regime.

	% ==============================

	% Loss-Cone Scattering -----
    \vspace{0.1in}
    \item
    \textbf{Stellar Loss-Cone (LC) scattering} ($\simclose \mathrm{pc}$), is implemented using the prescription from \citet{sesana2006} \& \citet{sesana2010} following the formalism of \citet{quinlan1996}.  Here, the hardening rate ($da/dt$) and eccentricity evolution ($de/dt$) are determined by dimensionless constants $H$ and $K$, calculated in numerical scattering experiments such that,
        \begin{equation}
    	\label{eq:dadt_lc_emp}
        \frac{da}{dt} \bigg|_{u} \equiv - \frac{G \rho}{\sigma}  a^2 \, H ,
        \end{equation}
and 
        \begin{equation}
    	\label{eq:dedt_lc_emp}
        \frac{de}{dt} \bigg|_{u} \equiv \frac{G \rho}{\sigma} a \, H \, K.
        \end{equation}
Here the binary separation (semi-major axis, $a$) and eccentricity ($e$) are evolved based on profiles of density ($\rho$) and velocity dispersion ($\sigma$) calculated from each binary host-galaxy in Illustris.  We use the fitting formulae and tabulated constants for $H$ and $K$ from \citet{sesana2006}\footnote{\citet{sesana2006}: Eqs.~16 \& 18, and Tables 1 \& 3, respectively}.  Note that this semi-empirical approach, which we will refer to as \textit{`eccentric LC models'}, explicitly assumes a full loss-cone in the scattering experiments by which they are calibrated.

	\vspace{0.1in}
    In our previous calculations presented in \citetalias{paper1}, we used a different LC prescription for binaries restricted to circular orbits.  In this paper we focus on the \textit{eccentric LC models}, but include results with our previous \textit{`circular'} prescription for comparison.  The circular models\footnote{
    These follow the theoretically-derived formulae from \citet{magorrian1999} for a spherically-symmetric background of stars scattering with a central object.  Scattering rates are calculated assuming isotropic, Maxwellian velocities and stellar distribution functions calculated from each galaxy's stellar density profile.  The density profiles are extended to unresolved ($\lesssim \mathrm{pc}$) scales with power-law extrapolations.  Additional details on Illustris stellar densities are included in \secref{sec:ap_gal_dens}.}
include a dimensionless `refilling parameter', ${\frefill \in \{0.0, 1.0\}}$, which interpolates between a `steady-state' LC (${\frefill = 0.0}$), where equilibrium is reached between the scattering rate and refilling by the two-body diffusion of stars; and a `full' LC ($\frefill = 1.0$), where the stellar distribution function is unaltered by the presence of the scattering source.  The $\frefill$ parameter for true astrophysical binaries is highly uncertain, but can drastically affect the efficiency with which binaries coalesce \citepalias[see the discussion in][]{paper1}.  We explore six models with, ${\frefill = [0.0,\, 0.2,\, 0.4,\, 0.6,\, 0.8,\, 1.0]}$.  Some recent studies tend to favor nearly-full LC models \citep[e.g.][]{sesana2015, vasiliev2015}.  The eccentric LC model, with imposed zero-eccentricities ($e_0 = 0.0$), yields results very similar to the circular-only model with a full LC ($\frefill = 1.0$), as expected.

	% Viscous Disk -----
    \vspace{0.1in}
    \item
    \textbf{Gas drag from a circumbinary, Viscous Disk (VD}; $\simclose 10^{-3} \, \mathrm{pc}$) is calculated following the thin disk models from \citet{hkm09}.  In these models, the disk is composed of three, physically distinct regions \citep{shapiro1986} determined by the dominant pressure (radiation versus thermal) and opacity (Thomson versus free-free) sources.  From inner- to outer- disk, the regions are: 1) radiation \& Thomson, 2) thermal \& Thomson, and 3) thermal \& free-free. The disk density profiles are constructed based on the self-consistently derived accretion rates given by Illustris, and truncated based on a (Toomre) gravitational stability criterion.  Higher densities resulting from higher accretion rates lead to more extended inner-disk regions.  We use an alpha-disk\footnote{Where viscosity depends on both gas and radiation pressure, as apposed to a `beta-disk' which depends only on the gas-pressure.  The differences in merger times and coalescing fractions between the alpha and beta models are negligible.  We use the alpha model because it may be more conservative via higher viscosities in the inner-most disk regions which, while insignificant for increasing the number of merging MBH, could increase GWB attenuation.} throughout.  It's worth noting that the inner-disk region (1) has a very similar hardening curve to GW-emission: ${\bigt_\tr{VD,1} \propto r^{7/2}}$, versus ${\thardgw \propto r^{4}}$.  In Illustris, post-merger MBH tend to have higher accretion rates, and thus larger inner-disk regions.
    
    \vspace{0.1in}
    We assume that the disk has a negligible effect on the eccentric
    evolution of binaries, i.e.~${\left[de/dt\right]_\textrm{VD} = 0}$.  This assumption is made for simplicity.
	While numerous studies have shown that eccentric evolution can at times be significant in circumbinary
	disks \citep[e.g.][]{armitage2005, cuadra2009, roedig2011}, we are unaware of generalized
	descriptions of eccentricity evolution for binaries/disks with arbitrary initial configurations.
	In the analysis which follows, we explore a wide range of eccentricity parameter space.  While a given
	model may end up being inconsistent with VD eccentric-evolution, the overall parameter space should
	still encompass the same resulting GWB spectra.

	% Gravitational Waves -----
    \vspace{0.1in}
    \item
    \textbf{Gravitational Wave (GW) emission} ($\simclose 10^{-5} \, \mathrm{pc}$) hardens binaries at a rate
 	given by \citet[][Eq.~5.6]{peters1964} as, 
        \begin{equation}
        \label{eq:dadt_gw}
    	\frac{da}{dt} = - \frac{64 \, G^3}{5 \, c^5} \frac{M_1 \, M_2 \left(M_1 + M_2\right)}{a^3} \, F(e),
        \end{equation}
    where the eccentric enhancement,
        \begin{equation}
        \label{eq:fe}
    	F(e) \equiv \frac{\left( 1 + \frac{73}{24} e^2 + \frac{37}{96} e^4 \right)}{\left(1 - e^2\right)^{7/2}}.
        \end{equation}
	In \citetalias{paper1} we made the approximation that the eccentricity of all binaries
	was negligible and thus ${F(e) = 1}$.  Here, we include models with non-zero eccentricity,
	evolved as \citep[][Eq.~5.7]{peters1964},
        \begin{equation}
        \label{eq:dedt_gw}
    	\frac{de}{dt} = - \frac{304 \, G^3}{15 \, c^5} \frac{M_1 \, M_2 \left(M_1 + M_2\right)}{a^4}
    					\frac{\left(e + \frac{121}{304} e^3\right)}{\left(1 - e^2\right)^{5/2}}. 
        \end{equation}

    \end{itemize}

% Gravitational Waves from Eccentric Binaries
% -------------------------------------------
\subsection{Gravitational Waves from Eccentric MBH Binaries}
\label{sec:meth_gw}
Circular binaries, with (rest-frame) orbital frequencies $\frest$, emit GW monochromatically at
$2 \frest$, i.e.~the $n=2$ harmonic.  Eccentric systems lose symmetry, and emit at $n=1$ and all
higher harmonics, i.e. $\fharm = n \, \frest$ (for $n \in \mathbb{I}$).  The GW energy spectrum
can then be expressed as \citep[][Eq.~3.10]{enoki2007a},
	\begin{equation}
	\label{eq:gw_energy_spectrum}
	\frac{d \egw}{d\frest} = \sum_{n=1}^\infty \left[ \lgwc(\fharm)
		\frac{\tfhard(\fharm, e)}{n \, \fharm} \, g(n,e) \right]_{\fharm = \frest/n}.
	\end{equation}
The GW frequency-distribution function $g(n,e)$ is shown in \refeq{eq:gw_freq_dist_func}.
Equation~\eqref{eq:gw_energy_spectrum} describes the GW spectrum
emitted by a binary \textit{over its lifetime}, which is used in the semi-analytic GWB
calculation (\secref{sec:gwb_sa}).
The total power radiated by an eccentric binary is enhanced by the factor $F(e)$, i.e.,
	$\lgw(\frest, e) = \lgwc(\frest) \cdot F(e)$,
where the GW luminosity for a circular binary is \citep[][Eq.~16]{peters1963},
	\begin{equation}
	\lgwc(\frest) = \frac{32}{5 G c^5} \left(G \mchirp \, 2 \pi \frest\right)^{10/3}.
	\end{equation}
Note that in \refeq{eq:gw_energy_spectrum}, the relevant timescale is the hardening-time
(or `residence'-time) \textit{in frequency},
	\begin{equation}
	\label{eq:time_hard}
	\tfhard \equiv \left| \frac{f}{df/dt} \right| = \frac{2}{3} \left| \frac{a}{da/dt} \right|
		\equiv \frac{2}{3} \thard,
	\end{equation}
which is $2/3$ the hardening-time \textit{in separation} (via Kepler's law), which we use for most of our
discussion and figures.

The GW strain from an individual, eccentric source can be related to that of a circular source as
\citep[e.g.][Eq.~9]{amaro-seoane2010}\footnote{Note the factor of $(2/n)^2$ when converting from
circular to eccentric systems.},
	\begin{equation}
	\label{eq:gw_strain_instant}
	\hs^2(\frest) = \sum_{n=1}^\infty \hsc^2(\fharm) \left(\frac{2}{n}\right)^2 \,
		g(n,e) \bigg|_{\fharm = \frest/n}.
	\end{equation}
Here, the GW strain from a circular binary is,
	\begin{equation}
	\label{eq:source_strain}
	\hsc(\frest) = \frac{8}{10^{1/2}} \frac{\left(G\mchirp\right)^{5/3}}{c^4 \, \lumdist}
		\left(2 \pi \frest\right)^{2/3}
	\end{equation}
\citep[e.g.][Eq.~8]{sesana2008}, for a luminosity distance $\lumdist$, and a chirp mass
$\mchirp = \left(M_1 M_2\right)^{3/5} / \left(M_1 + M_2\right)^{1/5}$.  Equation~\eqref{eq:gw_strain_instant}
describes the \textit{instantaneous} GW strain amplitude from a binary, and is used in the Monte-Carlo
GWB calculation (\secref{sec:gwb_mc}).

The GWB is usually calculated in one of two ways \citep{sesana2008}: either
Semi-Analytically (SA), treating the distribution of binaries as a smooth,
continuous and deterministic function to calculate
	$\partial^5 n_c(M_1, M_2, z, f_r, e)/ \partial M_1 \partial M_2 \partial z \partial f_r \partial e$
\citep{phinney2001},
or alternatively, in the Monte Carlo (MC) approach, where $n_c(M_1, M_2, z, f_r, e)$
is considered as a particular realization of a finite number of MBHB in the universe
\citep{rajagopal1994}.

% Semi-Analytic GWB
% -----------------
\subsubsection{Semi Analytic GWB}
\label{sec:gwb_sa}
The GWB spectrum can be calculated from a distribution of eccentric binaries as \citep[][Eq.~3.11]{huerta2015, enoki2007a},
	\begin{align}
	\label{eq:strain_sa_int}
	\begin{split}
	h_c^2(\fobs) = \frac{4G}{\pi c^2 \fobs} & \int dM_1 \, dM_2 \, dz \, n_c\left(M_1, M_2, z\right) \\
		& \sum_{n=1}^\infty \left[ \lgwc(\frest) \frac{\tfhard}{n \, \frest} \, g(n,e) \right]_{\frest = \fobs(1+z)/n},
	\end{split}
	\end{align}
where the summation is evaluated for all rest-frame frequencies with a harmonic matching the observed
(redshifted) frequency bin $\fobs$.  \refeq{eq:strain_sa_int} is derived by integrating the emission
of each binary over its lifetime, which is assumed to happen quickly ($\tfhard \ll \thubble$).

Each of the binaries in our simulation is evolved from their formation time (identified in Illustris) until
coalescence.  The GWB calculations only include the portions of the evolution which occur before redshift zero.
In our implementation of the \refeq{eq:strain_sa_int} calculation, interpolants are constructed for each
binary's parameters (e.g.~frequency, GW strain, etc) over its lifetime which are then used when sampling 
by simulated PTA.

% GWB: Monte Carlo
% ---------------
\subsubsection{Monte Carlo GWB}
\label{sec:gwb_mc}
The GWB spectrum can also be constructed as the sum of individual source strains for all binaries emitting
at the appropriate frequencies (and harmonics) in the observer's past light cone \citep[][Eq.~6~\&~10]{sesana2008},
	\begin{equation}
	\label{eq:strain_mc_int}
	\begin{split}
	\hc^2(f) = & \int dz \, d\mchirp \left[ \frac{d^3 N}{dz \, d\mchirp \, d \ln \frest} \, \hs^2(\frest) \right]_{\frest = \fobs(1+z)} \\
		= & \int dz \, d \mchirp \frac{d^2 n_c}{dz \, d\mchirp} d V_c \left[ \frac{\frest}{d \frest} \, \hs^2(\frest) \right]_{\frest = \fobs(1+z)},
	\end{split}
	\end{equation}
for a number of sources $N$, or comoving number-density $n_c$ in a comoving volume $V_c$.

The differential element of the past light cone can be expressed as \citep[e.g.][Eq.~28]{hogg1999},
	\begin{equation}
	dV_c(z) = 4\pi \left(1+z\right)^2 \frac{c}{H_0} \frac{d^2_c(z)}{E(z)} dz,
	\end{equation}
for a comoving distance $\comdist = \lumdist/(1+z)$, redshift-zero Hubble constant $H_0$, and the cosmological
evolution function $E(z)$ (given in \refeq{eq:cosmo_func}).  The term
$\frest/d\frest = \fobs / d\fobs$, which results from the \textit{definition} of the characteristic strain as
that over a logarithmic frequency interval, can be identified as the number of cycles each binary spends emitting
in a given frequency interval (see Eq.~\ref{eq:strain_from_sources}).

To discretize \refeq{eq:strain_mc_int} for a quantized number of sources (e.g.~from a simulation) we convert
the integral over number density, into a sum over sources within the Illustris comoving volume $\volill$,
	\begin{equation}
	\int dz \, d \mchirp \blank \frac{d^2 n_c}{dz \, d\mchirp} d V_c
		\rightarrow \sum_{ij} \blank \frac{\Delta V_{ij}}{\volill},
	\end{equation}
where the summation is over all binaries $i$ at each time-step $j$.
The factor $\frac{\Delta V_{ij}}{\volill} \equiv \volfactor$ represents
\textit{the number of MBH binaries in the past light cone represented by each binary in the
simulation}\footnote{$\volfactor$ is equivalent to the multiplicative factors used in, for example,
\citet[][Fig.~6]{sesana2008} and effectively the same as in \citet{mcwilliams2014}.}. 
The volume of the past light cone represented by $\volfactor$ depends on the integration step-size, i.e.,
	\begin{equation}
	\volfactor = \frac{1}{\volill}\frac{dV_c(z_{ij})}{dz_{ij}} \Delta z_{ij},
	\end{equation}
where $\Delta z_{ij}$ is the redshift step-size for binary $i$ at time step $j$.
$\volfactor$ is stochastic, determined by the number of binaries in a given region of the universe.
Alternative `realizations' of the universe can be constructed by, instead of using $\volfactor$ itself,
scaling by a factor drawn from a Poisson distribution $\poisson$, centered at $\volfactor$.  Thus, to construct
a particular realization of the GWB spectrum we calculate,
	\begin{equation}
	\label{eq:strain_mc_sum}
	\hc^2(f) = \sum_{ij} \poisson(\volfactor) \sum_{n=1}^\infty \left[ \frac{\frest}{\Delta f} \, \hs^2(\frest) \left(\frac{2}{n}\right)^2 g(n,e) \right]_{\frest = \fobs(1+z)/n}.
	\end{equation}

% Detection with PTA
% ------------------
\subsection{Detection with Pulsar Timing Arrays}
\label{sec:meth_pta}
To calculate the detectability of our predicted GWB spectra, we use the detection formalism outlined by \citet{rosado2015}.  A `detection statistic'\footnote{i.e.~measure of signal strength in PTA (mock) data.} $X$ is constructed as the cross-correlation of PTA data using a filter which maximizes the Detection Probability (DP) $\gamma$.  The optimal filter is known to be the `overlap reduction function' \citep{finn2009} which, for PTA, is the \citet{hellings1983} curve that depends on the particular PTA configuration (angular separation between each pair of pulsars).  Using the optimal detection statistic, and the noise characteristics of the PTA under consideration, parameters like the signal-to-noise ratio (SNR; and SNR-threshold) or DP can be calculated based on a GWB.  \citet{rosado2015} should be consulted for the details of the detection formalism but, for completeness, the relevant equations used in our calculations are included in \secref{sec:ap_det}.  

Following \citet{rosado2015} and \citet{taylor2016} we construct simulated PTA using published specifications of the constituent pulsars.  We then calculate the resulting DP (\refeq{eq:det_prob_b}) for our model GWB against each PTA, focusing on the varying time to detection.  We consider models for all PTA:

	\begin{itemize}
	\item
	European\footnote{\url{http://www.epta.eu.org/aom/EPTA_v2.2_git.tar}}
	\citep[EPTA,][]{desvignes2016, caballero2016}

	\item 
	NANOGrav\footnote{\url{https://data.nanograv.org/}}
	\citep[][]{arzoumanian2015a},
	
	\item
	Parkes\footnote{\url{http://doi.org/10.4225/08/561EFD72D0409}}
	\citep[PPTA,][]{shannon2015, reardon2016},
	
	\item
	International\footnote{\url{http://www.ipta4gw.org/?page_id=519}}
	\citep[IPTA,][]{verbiest2016, lentati2016}.

	\end{itemize}
	
For each array we include an `expanded' model (denoted by `+') including the addition of a pulsar every $X$ years, where for the individual PTA, $X = 1/4$, and for the IPTA, $X = 1/6$ \citep{taylor2016}.  All expanded pulsars are given a TOA accuracy of $250$ ns, and a random sky location.  The public PTA specifications include observation times for each TOA of all pulsars in the array.  We take the first TOA as the start time of observations for the corresponding pulsar, and use the overall number of calendar days with TOA measurements\footnote{Grouping TOA by observation day deals with near-simultaneous observations at different frequencies which are not representative of the true observing cadence.} to determine the characteristic observing cadence.  The cadence from the PTA data files is assumed to continue for the pulsars added in expansion.

Detection statistics depend on pairs of pulsars.  For each pair, we set the observational duration as the stretch of time over which both pulsars were being observed: $T_{ij} = T_i \cap T_j$; and take the characteristic cadence as the maximum from that of each pulsar: $\Delta t_{ij} = \max(\Delta t_i, \Delta t_j)$.  The sensitive frequencies for each pair is then determined by Nyquist sampling with $\Delta f_{ij} = 1 / T_{ij}$, such that each frequency $f_{ijk} = 1/T_{ij} + k/\Delta t_{ij}$, and $f \in [1/T, 1/\Delta t]$.  In calibrating the $\scalenoise$ parameter, we take the end time of observations as the last TOA recorded in the public data files, while for calculations of time to detection, we start with an end point of 2017/01/01. 

Pulsars are characterized by a (white-noise) standard deviation $\sigma_i$ in their TOA.
For a time interval $\Delta t$, the white-noise power spectrum $P_{w,i}$ is given by,
	\begin{equation}
	P_{w,i} = 2\sigma_i^2 \Delta t.
	\end{equation}
Some pulsars in the public PTA data provide specifications for a red-noise term\footnote{Models for each
PTA are: European--\citet[Eq.3]{caballero2016}; NANOGrav--\citet[Eq.4]{arzoumanian2015a};
Parkes--\citet[Eq.4]{reardon2016}; International--\citet[Eq.10]{lentati2016}.  Note that for the red-noise
amplitudes included in the IPTA public data release, the frequency $f$ must be given in $\pyr$, and the
duration $T$ in yr \citep[for Eq.10 of][]{lentati2016}.} which we also include.
We assume that when red-noise specifications are not provided that they are negligible.

To calibrate our calculations to the more comprehensive analyses employed by the PTA groups themselves,
we rescale the white noise ($\sigma_i$) of each pulsar by a factor $\scalenoise$ \citep[the procedure
described in][]{taylor2016}.
To determine $\scalenoise$, we calculate upper-limits on the GWB amplitude $\aul$ \citep[][Eq.~4]{taylor2016},
with a `true' (i.e.~injected) GWB amplitude of $\ayr = 0.6\E{-15}$, and iteratively adjust
$\scalenoise$ until the calculated $\aul$ matches the published values.  
The total noise used in our calculations is then\footnote{In the presence
of red-noise the power spectrum is frequency dependent, i.e.~$P_{ik} = P_i(f_k)$, but we suppress the
additional subscript for convenience.},
	\begin{equation}
	\label{eq:total_noise_power}
	P_{i} = \scalenoise^2 \, P_{w,i} + P_{r,i}.
	\end{equation}

Detailed specifications of each PTA configuration are included online as \texttt{JSON} files.  The basic
parameters of each individual array and the IPTA are summarized in \tabref{tab:pta}.
The values of $\scalenoise$ say something about how consistent the overall noise-parameters
are with the upper-limits calculated \textit{in our framework}.  Values of $\scalenoise > 1$ suggest that
additional noise is required.

    \begin{table}\centering
    \renewcommand{\arraystretch}{1.2}
    \setlength{\tabcolsep}{4pt}
    \begin{tabular}{@{}l cc cc cc@{}}
	    			 		&					&						& \multicolumn{3}{c}{Medians (Pulsars / Pairs)}		&					\\ 
	\cmidrule(lr){4-6}
    \multirow{2}{*}{Name}	& \multirow{2}{*}{N}& \multirow{2}{*}{Red} 	& $\sigma$ 			& Dur.				& Cad.		& \multirow{2}{*}{$\scalenoise$}	\\ 
  		 				 	& 					&						& [$\mu$s]			& [yr]		 		& [day]		& 				\\ \midrule
	European				& 42				& 8						& 6.5 / 6.9			& 9.7 / 8.2			& 14 / 20	& 2.26			\\
	NANOGrav				& 37				& 10					& 0.31 / 0.26		& 5.6 / 2.3			& 14 / 14	& 3.72			\\
	Parkes					& 20				& 15					& 1.8 / 1.8			& 15.4 / 9.1		& 21 / 23	& 0.1 			\\
	\iptazero				& 49				& 16					& 3.5 / 3.4			& 10.8 / 5.8		& 15 / 23	& 5.46			\\
	\iprime					& 49				& 27					& 1.2 / 1.6			& 12.8 / 8.2		& 14 / 17	& 1.0 			\\
    \end{tabular}
    \caption{Summary of parameters for the individual and International PTA used in our calculations.  The first and second columns give the number of pulsars (N) and the number which include a red-noise model (Red) in the official specifications.  The following three columns---the noise ($\sigma$), and observational duration \& cadence---are each given as median value for `pulsars/pulsar-pairs'.  Durations are those up to the end-time of each public data set (ranging from 2011 for Parkes, to 2015 for the EPTA).  The observing cadence is calculated based on the total number of days with TOA entries between the first and last recorded observations.  The \iptazero{} is based on the official IPTA data release while the \iprime{} is a manual combination of specifications from each individual PTA data release, without calibrating to any published upper-limit (i.e.~$\scalenoise \equiv 1.0$), but using the calibration from the individual PTA.}
    \label{tab:pta}
    \end{table}

The International PTA (\iptazero{}) requires the largest $\scalenoise$, suggesting that either the noise is under estimated or that the calculated upper-limit is sub-optimal---possibly due to systematics in combining data from not only numerous telescopes, but numerous groups and/or methodologies.  To address this issue we also present our results analyzed against an alternative ` \iprime{} '.  The \iprime{} is constructed by manually combining the TOA measurements from the individual arrays but without re-calibrating.  The total number of pulsars across all PTA is 68, but we only include those from the official IPTA specification (49) as the additional 19 pulsars produce a $\lesssim 1\%$ improvement in the resulting statistics.

To construct the \iprime{}, white noise parameters are added in quadrature using the $\scalenoise$ from each individual PTA.  When a pulsar has multiple red-noise models from different groups, we use the model with the lowest noise power at $f = 0.1 \, \pyr$.  The red-noise characteristics of the pulsars have a substantial effect on the resulting detection statistics.  Because the \iprime{} model incorporates the red-noise from all PTA, it ends up being significantly disadvantaged compared to the EPTA (for example) for which only 8 pulsars have red-noise models.  To level the playing field, we copy the \iprime{} red-noise models to pulsars of each individual PTA which do not otherwise include one.  Thus the actual number of pulsars using red-noise models, differs from the values shown in \tabref{tab:pta}, specifically: our EPTA, NANOGrav, and Parkes models end up with 21, 16, and 18 pulsars with red-noise, and both IPTA models have 27.

Our models for the expanded versions of the individual PTA and \iptazero{} begin adding pulsars after the end of their public data sets, which end in 2015/01 (EPTA), 2013/11 (NANOGrav), 2011/02 (PPTA), \& 2014/11 (\iptazero{}).  The parameters of the expansion pulsars are low white-noise and zero red-noise, which may give an unfair advantage, for example, to the PPTA vs.~the EPTA by adding roughly 16 low-noise pulsars by 2015/01.  To try to take this into account for the \iprime{}+ model, we add 2 pulsars per year after 2011/02 (when the PPTA data set ends), 4 per year after 2013/11, and finally 6 per year after 2015/01---the same time at which \iptazero{} begins expansion with 6 pulsars per year.

To test our PTA configurations and detection statistics, we reproduce the results of \citet[][Fig.~2]{taylor2016} for purely power-law GWB spectra.  Our detection probabilities for each PTA are shown in \figref{fig:dp_ppl}.  The left column shows expanded (+) arrays, with new pulsars added each year, while the right column shows the current array configurations.  Our Detection Probabilities (DP) are close to those of \citet{taylor2016}, but not identical, likely due to differences between our noise scaling-factors $\lambda_\tr{noise}$\footnote{
\citet{taylor2016} did not publish their $\scalenoise$, but they were calculated by marginalizing against a probability distribution for $\ayr$ from \citet{sesana2013}.  We simply use a fixed value based on the GWB from our fiducial MBHB merger model, ${\ayr = 0.6\E{-15}}$.}.
Each panel has vertical lines denoting the year in which each PTA reaches $50\%$ (short-dashed) and $95\%$ DP (long-dashed).  For these power-law GWB models at our fiducial ${\ayr = 0.6\E{-15}}$ \citepalias{paper1}, the individual PTA reach $95\%$ DP in roughly 2031, 2027 \& 2027 for the EPTA+, NANOGrav+ \& PPTA+ respectively.  The \iptazero{}+ reaches the same DP at 2029, while \iprime{}+ cuts that down to 2025.  \iptazero{}+ performing worse than NANOGrav and Parkes further motivates the usage of the \iprime{}+ model.  The difference between static and expanding arrays is quite significant.  For example, by 2037, none of the static PTA models reach $95\%$ DP.  This highlights the importance of continuing to observe known pulsars, while also surveying the sky looking for new ones.  Survey programs have been carried out by several large radio telescopes around the world, including GBT and Arecibo in the US, and their continued efforts and funding is of critical importance for PTA science.

% Create a bibliography here, only if just this file is being compiled/built.
\biblio{}

% Results
% ----------

% =============================================================
% =======================   RESULTS   =========================

\section{Results}
\label{sec:results}

% Eccentric Evolution
% -------------------
\subsection{Eccentric Evolution}
\label{sec:res_ecc}

	% ====    FIGURE 1 : Eccentricity Evolution    ====
    \begin{figure}
    \centering
	\includegraphics[width=\columnwidth]{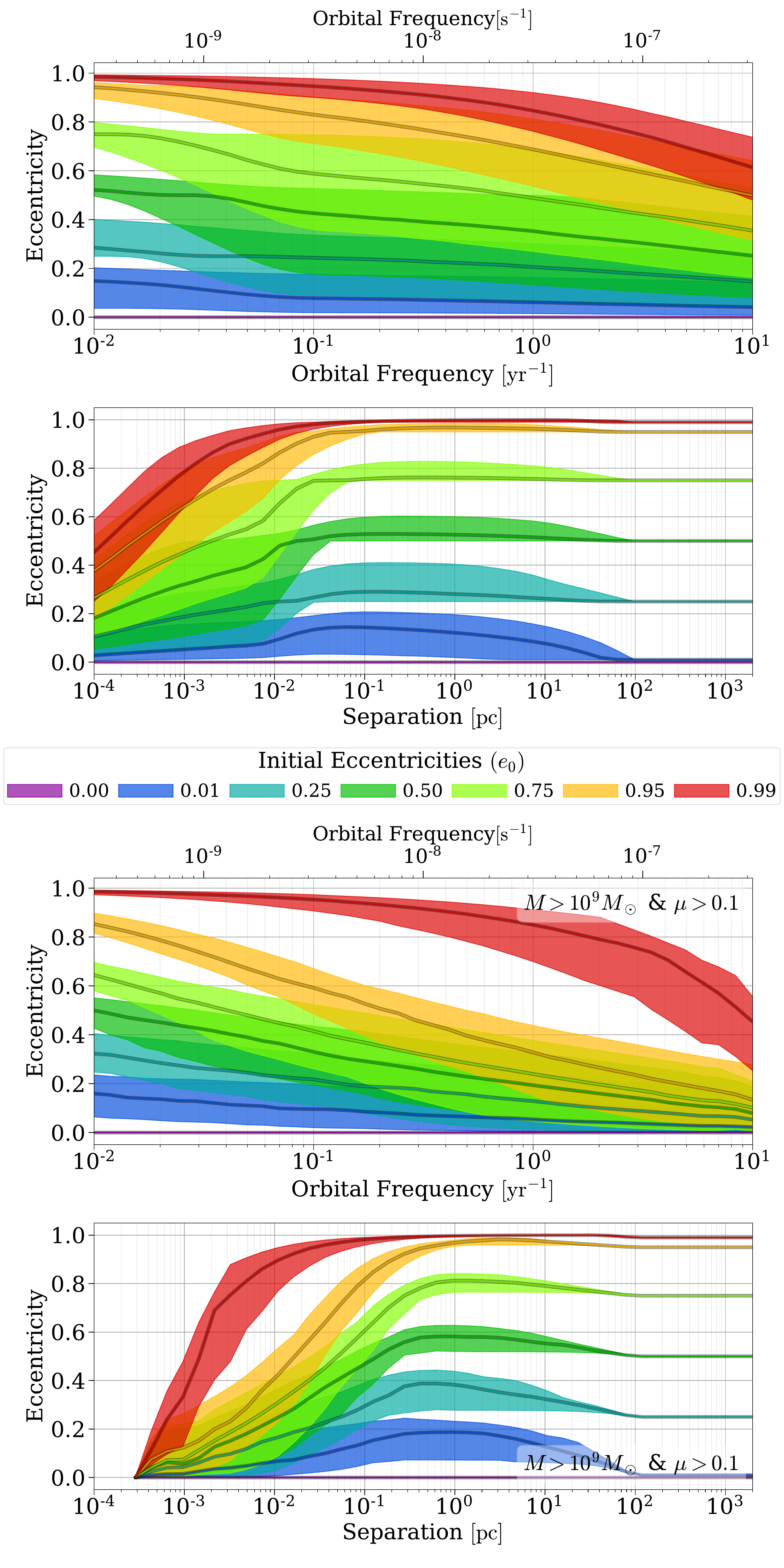}
    \caption{Evolution of eccentricity versus separation (panels a \& c) and orbital
    frequency (panels b \& d) for a variety of initial eccentricities.  The upper panels (a \& b)
    show eccentricities for all binaries, while the lower panels (c \& d) show only the \heavydef{}
    and \majordef{} subset which tend to dominate the GWB.  Each band corresponds to
    $68\%$ of the population, and the central lines to the medians.}
    \label{fig:eccen_evo}
    \end{figure}

%\FloatBarrier

% Eccentricity evolution
For a given simulation, all binaries are initialized with the same
eccentricity and are let to evolve with dynamical friction (DF), loss-cone (LC) stellar scattering,
drag from a circumbinary Viscous-Disk (VD), and Gravitational Wave (GW) hardening.  LC
increases initially-nonzero eccentricities, and GW emission decreases them.  In our models we assume DF \&
VD do not affect the eccentricity distribution.  The upper panels of \figref{fig:eccen_evo} show
the resulting binary eccentricity evolution versus orbital frequency (a) and binary separation
(b) for our entire population of MBHB.  Solid lines show median values at each separation, and
colored bands show the surrounding $68\%$.  Binaries initialized to ${e_0 = 0.0}$ stay at zero, but
the population initialized to only $e_0 = 0.01$ have a median roughly ten times larger at
${r \sim 1 \, \pc}$, and still ${e \gtrsim 0.05}$ at orbital frequencies of $1 \, \pyr$.

Binaries which are both \heavydef{} and \majordef{} tend to dominate the GWB signal \citep{paper1}.
The eccentricity evolution of this subset of binaries is shown separately in panels (c) \& (d) of
\figref{fig:eccen_evo}.  The eccentricities of these systems tend to dampen more quickly at
separations below $\simclose 1 \, \pc$.  In general, this leads to much lower eccentricities at PTA frequencies
for the \heavymajor{} population.  In the $e_0 = 0.95$ model, for example, the \heavymajor{} median
eccentricity drops below $0.6$ by ${f = 0.1 \, \pyr}$ (panel c), whereas it takes until ${f \sim 3 \, \pyr}$
for the population of all binaries (panel a).  In terms of separation, most of the ${e_0 = 0.95}$ population
has ${e \lesssim 0.5}$ by ${r \sim 10^{-2} \, \pc}$, whereas for all binaries the same isn't true until
${r \sim 10^{-4} \, \pc}$.  The highest eccentricity model, ${e_0 = 0.99}$, behaves somewhat differently,
as it tends to in many respects.  For ${e_0 = 0.99}$, binaries pass through most of the
PTA band before their eccentricities are substantially damped in both the overall and \heavymajor{}
groups.  

Once binaries begin to approach the PTA sensitive band (${f \gtrsim 10^{-2} \, \pyr}$) the eccentricity
distributions are always monotonically decreasing.  At the corresponding separations, GW becomes more
and more dominant to LC, but at the same time, VD can still be an important hardening mechanism.
If circumbinary disks can be effective at increasing eccentricity, or if a resonant
third MBH were present, eccentricities could still be excited in the PTA regime.  Either of these effects
may be important for some MBHB systems, but likely not for the overall population.

	% ====    FIGURE 2 : Hardening Timescale Evolution    ====
    \begin{figure}
    \centering
    \includegraphics[width=\columnwidth]{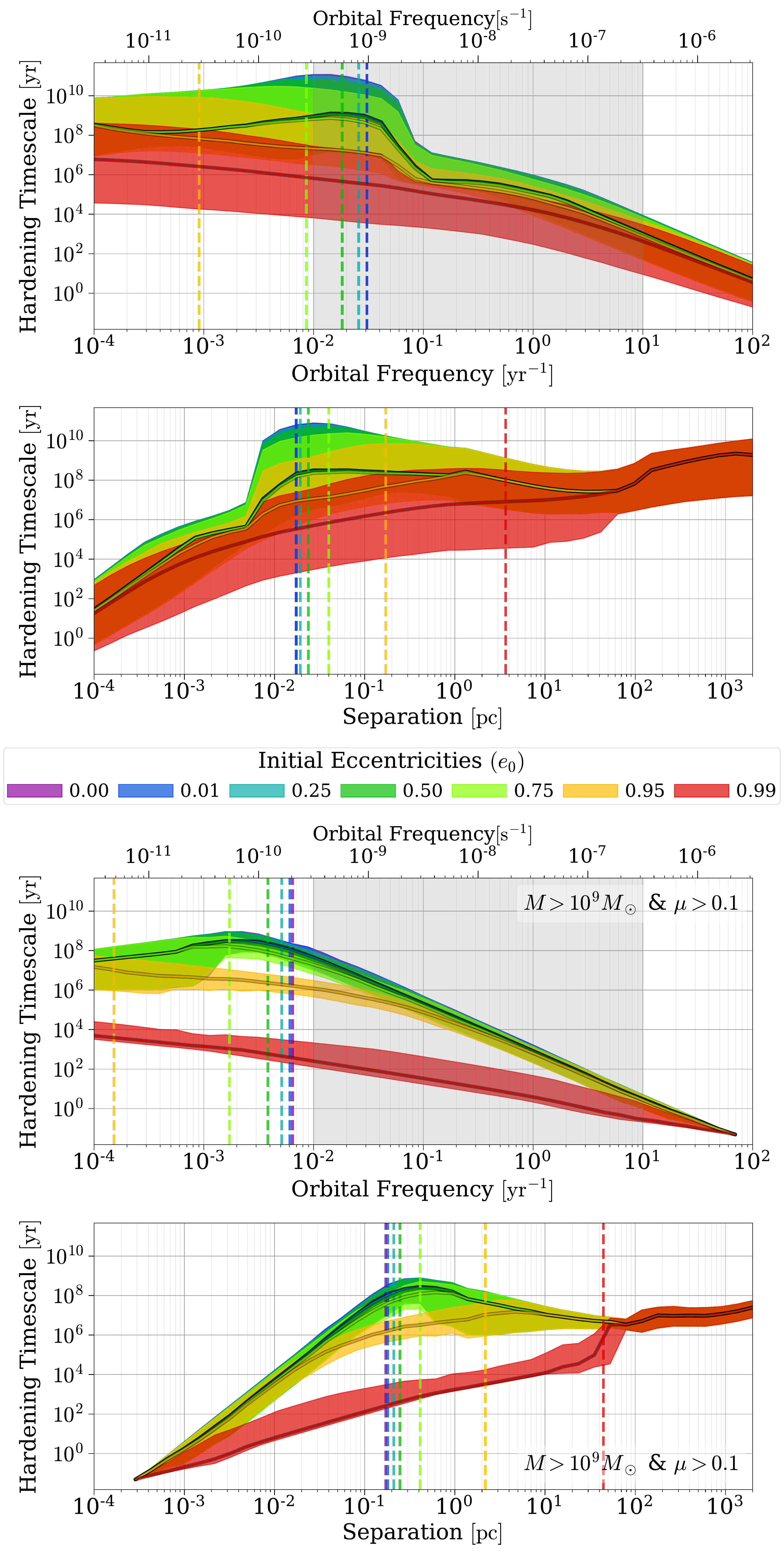}
    \caption{Binary hardening timescale ($a/(da/dt)$) against orbital frequency (panels a \& c) and binary separation (panels b \& d).  Note that the frequency axes show an extended range, with the PTA-sensitive band shaded in grey.  The upper panels (a \& b) show hardening rates for all binaries, while the lower panels (c \& d) show only the \heavymajor{} subset.  The dashed vertical lines indicate the radii at which GW-driven hardening becomes dominant over DF and LC.}
    \label{fig:eccen_hard}
    \end{figure}

% Hardening Rates
In our models, while binary eccentricity is evolved by both LC and GW hardening, the eccentricity
distribution itself only affects the rate of \textit{semi-major-axis} hardening (i.e.~$da/dt$,
\refeq{eq:dadt_gw}) by the GW mechanism.  As eccentricity increases, GW hardening becomes more
effective.  The hardening time ($a / [da/dt]$) for all binaries is plotted in the upper panels of
\figref{fig:eccen_hard}.  The frequency panel (a) has been extended to show more of the physical
picture, with the PTA-relevant regime shaded in grey.  The DF regime goes from large separations
down to ${r \sim 50 \, \mathrm{pc}}$, and according to our prescription includes no eccentricity
dependence.  

The relatively flat portion of the hardening curve between ${r \sim 50 \, \mathrm{pc}}$ and ${r \sim 2\E{-2} \, \mathrm{pc}}$ (panel b) is typically the LC-dominated regime.  Ignoring the radial dependence of galactic $\rho/\sigma$ (density/velocity) profiles, the hardening rate should scale like ${\thard \propto a^{-1}}$ (see \refeq{eq:dadt_lc_emp}).  The hardening rates in panel b, however, show the combined binary evolutionary tracks over many orders of magnitude in total-mass and mass-ratio, flattening the hardening curves.  The scaling is more clear for the low-eccentricity models of the \heavymajor{} subset (panel d) which show more canonical LC hardening rates between ${r\sim 10 \, \pc}$ and ${r \sim 1 \, \pc}$.

Even at these large separations, the hardening timescale is significantly decreased for the highest
initial eccentricities: ${e_0 = 0.95}$ and especially ${e_0 = 0.99}$.  In these models,
GW emission begins to play an important role in hardening much earlier in the systems' evolution.
The dashed vertical lines in \figref{fig:eccen_hard} show the frequency and separation at which
GW hardening becomes dominant over DF and LC\footnote{Note that VD can be dominant for an additional
decade of frequency or separation, but because the amplitude and power-law index of VD are very
similar to that of GW hardening, the transition points plotted are more representative of a change
in hardening rate and/or GWB spectral shape.} in $50\%$ of systems.  For ${e_0 = 0.99}$, GW domination
occurs at a few parsecs, largely circumventing the LC regime entirely.  As eccentricity decreases,
so does the transition separation.  For lower initial eccentricities, ${e_0 \lesssim 0.75}$,
the DF \& LC become sub-dominant at a few times ${10^{-2} \, \mathrm{pc}}$, and the overall hardening
rates are hardly distinguishable between different eccentricities.

Panels (c) \& (d) of \figref{fig:eccen_hard} show the hardening time for the population of
\heavymajor{} binaries.  While very similar to the overall population, the \heavymajor{} binaries
are all effectively in the GW regime at frequencies ${f \gtrsim 10^{-2} \, \pyr}$, i.e.~the entire
PTA band.  In the ${e_0 = 0.0}$ model, binaries in the PTA-band show a nearly perfect power-law
hardening rate from GW-evolution, and eccentricities ${e_0 \lesssim 0.75}$ are hardly different.
In the ${e_0 = 0.95}$ model, a heightened hardening rate is clearly apparent up to
${f \sim 10^{-1} \, \pyr}$.  The ${e_0 = 0.99}$ model evolves five orders of magnitude faster than
${e_0 = 0.0}$ at ${f = 10^{-2} \, \pyr}$, and still more than two orders faster even at ${f = 1 \, \pyr}$.
The ${e_0=0.99}$ model hardens drastically faster than the others both because of the strong
eccentricity dependence of the GW hardening rate (\refeq{eq:dadt_gw}) and because the eccentricity
of the ${e_0 = 0.99}$ population better retains its high values at smaller separations.

At frequencies below $10^{-3}$--$10^{-2} \, \pyr$ and lower eccentricity models (${e_0 \lesssim 0.75}$),
the \heavymajor{} population is LC-dominated, causing sharp turnovers in the their hardening curves
which is echoed in the resulting GWB spectra.  In the population of all binaries (panel a) the
transition occurs at slightly higher frequencies.  The higher eccentricity models, which transition
to GW domination at lower frequencies, do not show the same break in their hardening rate evolution.
The resulting GWB spectra, however, still do (discussed in \secref{sec:res_gwb}).

In all of our models, GW hardening is dominant to both dynamical friction and stellar scattering
for the binaries which will dominate GWB production (\heavymajor{} at ${f \gtrsim 10^{-2} \, \pyr}$).
Viscous drag, however, tends to be comparable or higher in much of the binary population.
Deviations to the power-law index of hardening rates in the PTA regime are not apparent, however,
because VD has a very similar radial dependence to GW emission\footnote{This is true for the
inner, radiation-dominated, region of the disk where GWB-dominating binaries tend to reside.}
(see \secref{sec:meth_eccen}).
Although the GWB spectrum is only subtly affected by VD, the presence of significant gas accretion
in the PTA band is promising for observing electromagnetic counterparts to GW sources, and
even multi-messenger detections with `deterministic'/`continuous' GW-sources---those MBHB resolvable
above the stochastic GWB \citep[e.g.][]{sesana2012, tanaka2012, spolaor2013}.

It should be clear from \figref{fig:eccen_hard} that the hardening timescale is often very long
(${\thard \sim \thubble}$) and varies significantly between binary systems.  A substantial fraction
of binaries, especially those with lower total mass and more extreme mass ratios, are unable to
coalesce before redshift zero.  The presence of significant eccentricity at sub-parsec scales can
significantly decrease the hardening timescale, increasing the fraction of coalescing systems overall.
The \heavymajor{} subset of binaries, however, already tends to coalesce more effectively, and because
this portion of the population dominates the GWB, its amplitude due to varying coalescing fractions
is only subtly altered.

%\FloatBarrier

% Gravitational Wave Backgrounds
% ------------------------------
\subsection{Gravitational Wave Backgrounds}
\label{sec:res_gwb}

	% ====    FIGURE 3 : GWB Spectra - Environmental vs. Eccentric Effects    ====
    \begin{figure}
    \centering
        \includegraphics[width=\columnwidth]{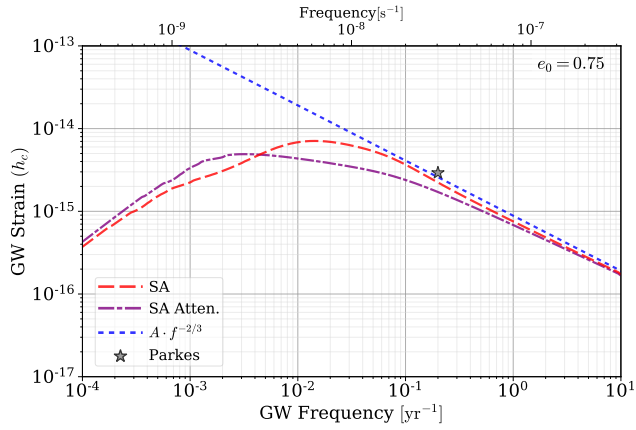}
    \caption{Stochastic Gravitational Wave Background calculated with the Semi-Analytic (SA) method, showing the effects of large binary eccentricities on the GWB spectrum.  A purely power-law model is shown in the blue dotted line, which includes the masses and redshifts of merger events but assumes all energy is emitted in GW.  The purple dot-dashed line includes the effect of attenuation from environmental hardening effects.  The red dashed line shows the full SA calculation including both attenuation and the redistribution of GW energy across multiple frequencies.  Note that this model uses a large initial eccentricity of $e_0 = 0.75$.}
    \label{fig:gwb_sa_eccen_effects}
    \end{figure}

% Smooth GWB (Semi-Analytic)
% - - - - - - - - - - - - -
\subsubsection{The Semi-Analytic GWB}
The simplest calculation of the GWB assumes that all binaries coalesce effectively, quickly,
and purely due to GW emission.  This leads to a purely power-law strain spectrum,
$h_\trt{c, GW} \propto f^{-2/3}$.  Environmental hardening, however, is required for the vast
majority of binaries to be able to coalesce within a Hubble time\footnote{Even with strong external
factors,  hardening timescales are still often comparable to a Hubble time, and thus only a fraction
of systems coalesce.}.  These additional hardening mechanisms also force binaries to pass
through each frequency band faster than they otherwise would, decreasing the GW energy
emitted in that band, and attenuating the GW background spectrum
\citep[e.g.~\refeq{eq:strain_sa_int}, or][]{ravi2014}.

Non-zero eccentricity \textit{increases} the instantaneous GW luminosity and decreases the hardening
time (\refeq{eq:dadt_gw}).  Also, the GW emission from eccentric binaries is no longer produced
monochromatically at twice the orbital frequency, but instead is redistributed, primarily from
lower to higher frequencies (\refeq{eq:gw_energy_spectrum}).  The effects of increased attenuation
and the frequency redistribution are shown separately in \figref{fig:gwb_sa_eccen_effects} for
${e_0 = 0.75}$.
The blue, dotted line shows a power-law spectrum (i.e.~GW-only); while the purple, dashed-dotted 
line includes attenuation from the surrounding medium (predominantly LC-scattering).  The red, dashed
line includes the effects of chromatic GW emission, in addition to attenuation, which shifts the
spectrum to higher frequencies.  This is the full, semi-analytic (SA; \secref{sec:gwb_sa}) calculation.
The spectral turnover is produced by the environmental attenuation, while the frequency redistribution
slightly increases the amplitude in the PTA-regime (${f \gtrsim 10^{-2} \, \pyr}$).

	% ====    FIGURE 4 : GWB Spectra - Eccentricity with Semi-Analytic Models    ====
    \begin{figure}
    \centering
    \includegraphics[width=\columnwidth]{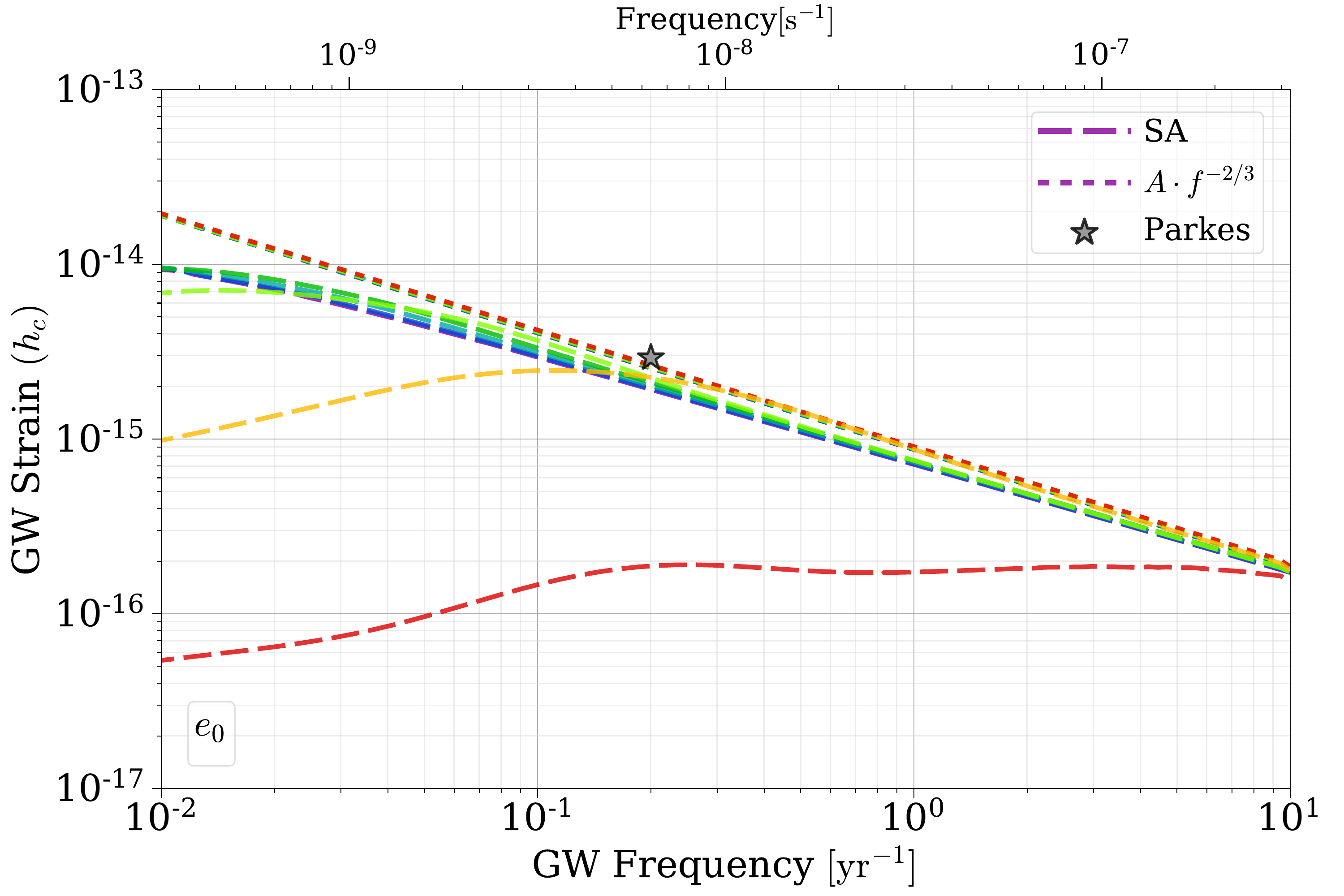}
    \caption{GWB strain spectrum calculate using the Semi-Analytic (SA) formalism for simulations with a range of initial eccentricities ($e_0$).  The dotted lines assume purely GW-driven evolution and no frequency redistribution, and show amplitude increasing slightly with increasing $e_0$---because more systems coalesce.  The dashed lines show the full SA calculation, including both attenuation and the frequency redistribution.  Note that the power-laws (dotted) are not physically meaningful \textit{per se}, and are meant as a reference when interpreting the SA models (dashed).  For ${f \gtrsim 2\E{-1} \, \pyr}$, the GWB amplitude increases with increasing $e_0$ until ${e_0 = 0.99}$, at which point the GWB amplitude is drastically depressed---almost completely flat---across the PTA band.}
    \label{fig:gwb_sa_eccen}
    \end{figure}

The GWB strain spectrum is shown for a variety of initial eccentricities in \figref{fig:gwb_sa_eccen}.
Purely power-law spectra calculated assuming GW-only evolution are shown in dotted lines, while the
full SA calculation is shown with dashed lines.  The power-law spectra show a very slightly increasing
GWB amplitude with increasing eccentricity as more binaries are able to coalesce before redshift zero.
For initial eccentricities ${e_0 \lesssim 0.5}$, the GWB spectrum above frequencies ${f \gtrsim 10^{-2} \, \pyr}$
are nearly identical.  For higher initial values, the effects of non-zero eccentricity become more apparent.
Overall, for all but the highest eccentricities ($e_0 > 0.9$), the effect of non-circular orbits is
actually to slightly increase the GWB amplitude in the PTA frequency band, because of the redistribution of GW
energy to higher harmonics.  Note that this is because the spectral turnover is always below the
PTA-band in our spectra \citep[also seen in, e.g.,][]{chen2016}.  If galactic densities are significantly
larger than those found in the Illustris
simulations, the environment could continue dominating binary evolution to higher frequencies.  If that
were the case, the spectral turnover could be within the PTA band which would lead to chromatic GW emission
\textit{decreasing} the observable signal.  In appendix \secref{sec:ap_gal_dens}, we include additional
information on binary host-galaxy densities from Illustris.

In the highest eccentricity case (${e_0 = 0.99}$; red lines), binaries are still highly eccentric at
separations corresponding to frequencies all the way up to ${f \sim 10 \, \pyr}$, and the GWB
amplitude is drastically diminished throughout the PTA band.  While such high eccentricities
may be unlikely \citep[e.g.][]{armitage2005, roedig2011}, having some subset of the population maintain
high eccentricities into the PTA band is certainly not impossible \citep[e.g.][]{rantala2016}.
If Kozai-Lidov-like processes from a hierarchical, third MBH were driving the eccentricity
\citep[e.g.][]{hoffman2007, bonetti2016}, or the binary were counter-rotating
to the stellar core or circumbinary disk \citep[e.g.][]{sesana2011, amaro-seoane2016}, eccentricities could
grow much faster than in our results.  Better understanding what fraction of systems could be susceptible
to these processes is an important direction of future study.

	% ====    FIGURE 5 : GWB Spectra - GWB by Mass Contribution    ====
    \begin{figure}
    \centering
    \includegraphics[width=\columnwidth]{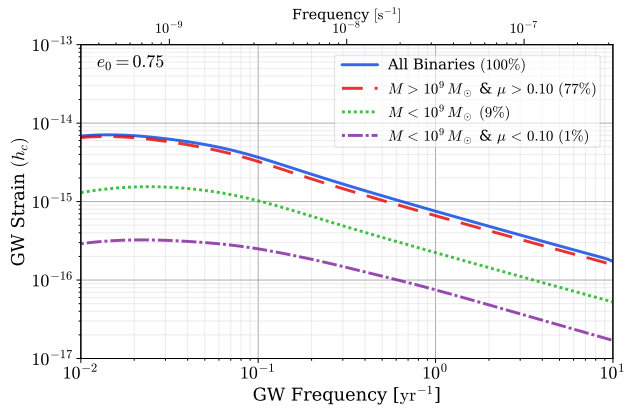}
    \caption{GWB from different binary subsets selected by total mass $M$, and mass ratio $\mu$.  The blue (solid)
    line shows the total GWB from all binaries, while the red (dashed) line shows the contribution from
    \heavymajor{} systems (shown also in Figs.~\ref{fig:eccen_evo}~\&~\ref{fig:eccen_hard}).  Light systems
    (green, dotted) and light \& minor binaries (purple, dot-dashed) are also shown.  The percentages in the
    legend show the contribution to the GWB energy at $1 \, \pyr$ (i.e.~${\propto \hc^2(f = 1 \, \pyr)}$) for
    each subgroup.  Here, the \heavymajor{} subgroup, while only $1\%$ of binaries by number, contributes almost
    $80\%$ of the GWB energy.}
    \label{fig:gwb_by_mass}
    \end{figure}

The contribution the GWB is broken down by subgroups of total mass and mass ratio in \figref{fig:gwb_by_mass}.  The complete GWB is shown in blue (solid), while the \heavymajor{} subset is in red (dashed), the light (${M < 10^9 \, \msol}$) in green (dotted), and the light \& minor (${\mu < 0.1}$) in purple (dash-dotted).  The \heavymajor{} binaries are only $\sim 1\%$ of the population, but here, in the $e_0 = 0.75$ model, contribute $\sim 80\%$ of the GWB energy density at $1 \, \pyr$.  In fact, \heavymajor{} systems dominate at all frequencies with a very similar spectral shape to the overall GWB, with a slight enhancement of GW strain at lower frequencies.  Lower-mass binaries ($M < 10^9 \, \msol$) contribute less than $10\%$ of the energy (despite being just over $90\%$ of the population), and systems which are neither heavy nor massive contribute only $1\%$ ($\sim 30\%$ of the population).  Other initial eccentricity models tend to be even more \heavymajor{} dominated (often $\sim 90\%$).

% Rough GWB (Monte-Carlo)
% - - - - - - - - - - - -
\subsubsection{The Monte-Carlo GWB}

	% ====    FIGURE 6 : GWB Spectra - Monte Carlo GWB for e_0 = 0.5    ====
    \begin{figure}
    \centering
    \includegraphics[width=\columnwidth]{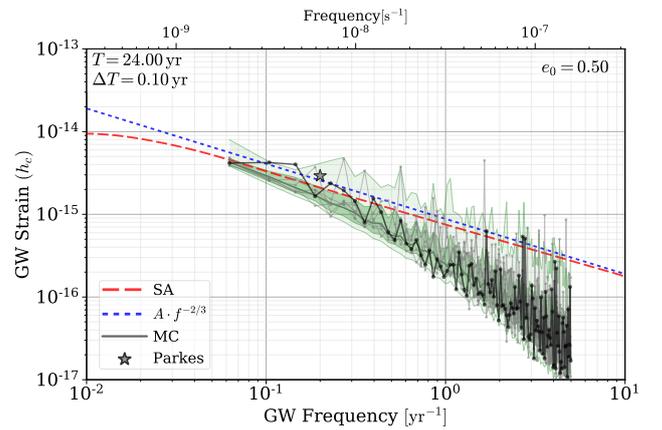}
    \caption{GWB calculated using the Monte-Carlo (MC) method, for a simulation with initial eccentricities ${e_0 = 0.5}$.  Seven different GWB realizations are shown in gray and black lines, while the median and one- \& two- sigma contours are shown in green.  Semi-Analytic (SA) calculations are also shown: purely power-law in blue (dotted), and full in red (dashed).  The high frequency spectrum of the MC calculation is sharper than that of the SA methods due to quantization of binaries.}
    \label{fig:eccen-0.5_gwb_mc}
    \end{figure}

The Monte-Carlo (MC) approach (\secref{sec:gwb_mc}) to calculating the GWB dispenses with the continuum
approximation for the density of GW sources, thereby allowing for the quantization of MBHB at the same
time as providing a convenient formalism for constructing an arbitrary number of realizations from the same
binary population.
The MC GWB is shown in \figref{fig:eccen-0.5_gwb_mc}, with seven randomly-chosen realizations plotted in
gray and black.  The median line and one- \& two- sigma contours of 200 realizations are
shown in green.  For reference, the SA spectra are shown for purely power-law calculations (blue, dotted) and
the full SA (red, dashed).  The frequency bins correspond to Nyquist
sampling at a cadence $\Delta T = 0.1 \, \mathrm{yr}$, and total observational-duration
$T = 24 \, \mathrm{yr}$.

The MC GWB differs from the SA one, both in its jaggedness and also in a steeper spectrum at higher frequencies.
The jaggedness is caused by varying numbers of binaries in the observer's past light cone
(\refeq{eq:strain_mc_sum})---especially massive ones at lower redshifts.  The latter effect is due to
binary quantization: at high frequencies, where the hardening time is short, there are few MBHB contributing
to each frequency bin.  The SA calculation, however, implicitly includes the contribution from fractional
binaries, artificially inflating the GWB amplitude at high frequencies \citep{sesana2008}.

	% ====    FIGURE 7 : GWB Spectra - Eccentricity with Monte Carlo    ====
    \begin{figure}
    \centering
    \includegraphics[width=\columnwidth]{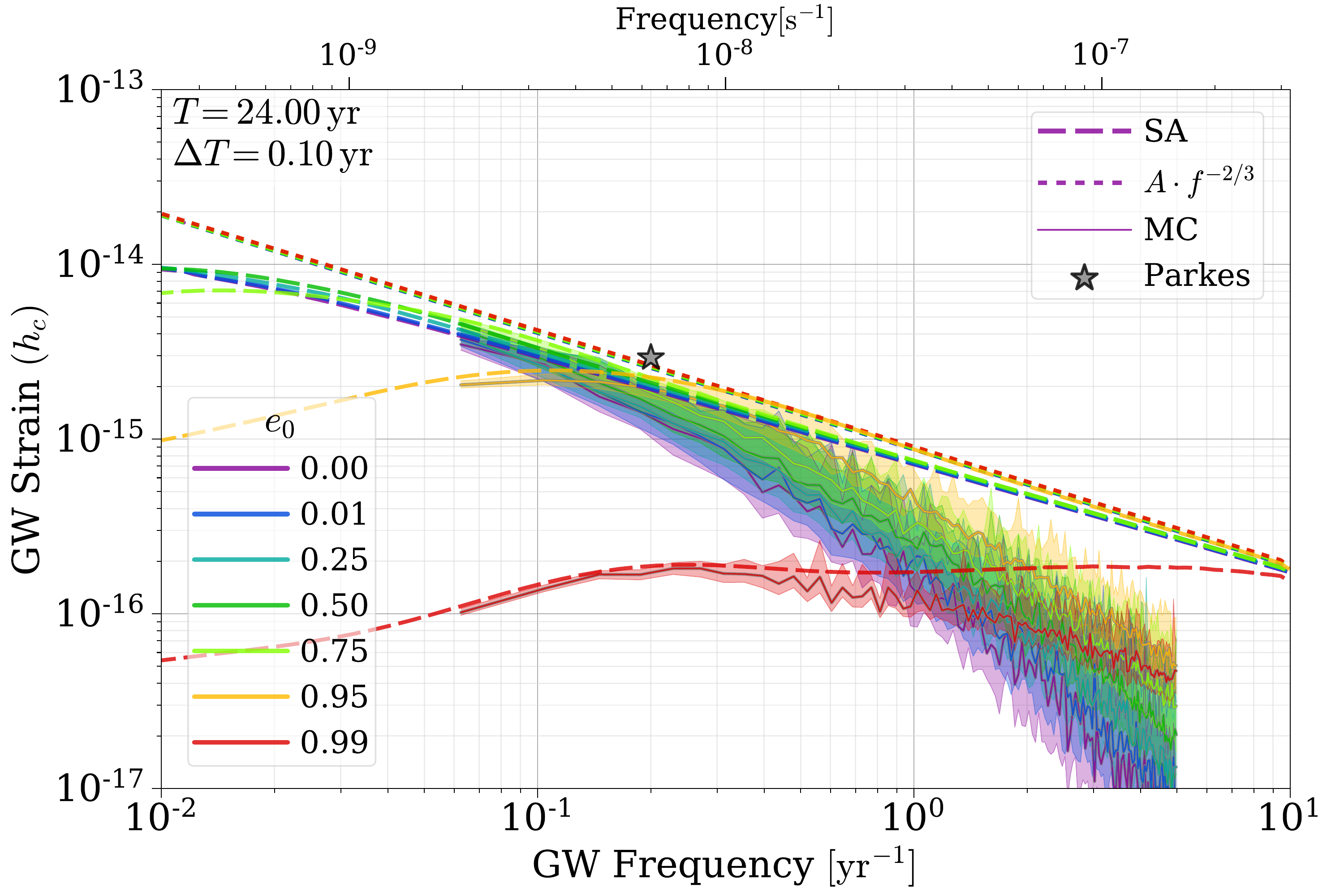}
    \caption{Monte-Carlo (MC) calculated GWB for a variety of different initial eccentricities ($e_0$).
    Median lines and one-sigma contours are shown for the MC case, and both the full (dashed) and power-law
    (dotted) SA calculations are also shown.  Increasing GWB amplitude with increasing eccentricity is
    apparent at lower frequencies (${\sim 10^{-1} \, \pyr}$, until the spectrum turns over.  While the
    spectral turnover is produced by environmental interaction, the frequency at which it occurs is
    increased with rising eccentricity.  At higher frequencies (${\gtrsim 2\E{-1} \, \pyr}$), as
    eccentricity increases, the MC results come closer and closer to the SA ones due to additional binaries
    at lower \textit{orbital} frequencies contribute to higher \textit{GW-frequency} bins.}
    \label{fig:gwb_mc_eccen}
    \end{figure}

Figure~\ref{fig:gwb_mc_eccen} shows the GWB for a variety of initial eccentricities, with median lines and
one-sigma contours for the MC calculation, along with the full and power-law SA calculations.  Both the SA
and MC methods show slightly increased GWB amplitudes with increasing eccentricities (except for the highest,
$e_0 = 0.99$ simulation).  A more pronounced effect is that the higher the eccentricity, the closer the
high-frequency portion of the MC calculation comes to the purely power-law spectra. The ratio of the MC GWB
(median lines) to SA GWB is shown explicitly in \figref{fig:gwb_mc_sa_rat}.  Higher eccentricities mean that
GW energy from binaries at lower \textit{orbital-frequencies} are deposited in higher \textit{GW-frequency}
bins.  This means that overall, more binaries are contributing to each of the higher frequency bins,
reducing the effect of MBHB quantization, and thus bringing the MC results closer in line to the SA
models.  Finite-number effects at high frequencies are also remediated by increased numbers of coalescing
systems, as is the case for very large LC refilling fractions ($\frefill$), for example, shown in
Figs.~\ref{fig:gwb_mc_frefill}~\&~\ref{fig:gwb_mc_sa_rat_frefill}.

% Pulsar Timing Array Detections
% ==============================
\subsection{Pulsar Timing Array Detections}
\label{sec:res_pta}

% DP: ECCENTRIC MODELS
% - - - - - - - - - - - -
\subsubsection{Eccentric Binary Evolution}

	% ====    FIGURE 8  :  eccentric DP vs. Time for 5 PTA   ====
	\newcommand{\dpscaleevo}{0.95}
    \begin{figure}
    \includegraphics[width=\dpscaleevo\columnwidth]{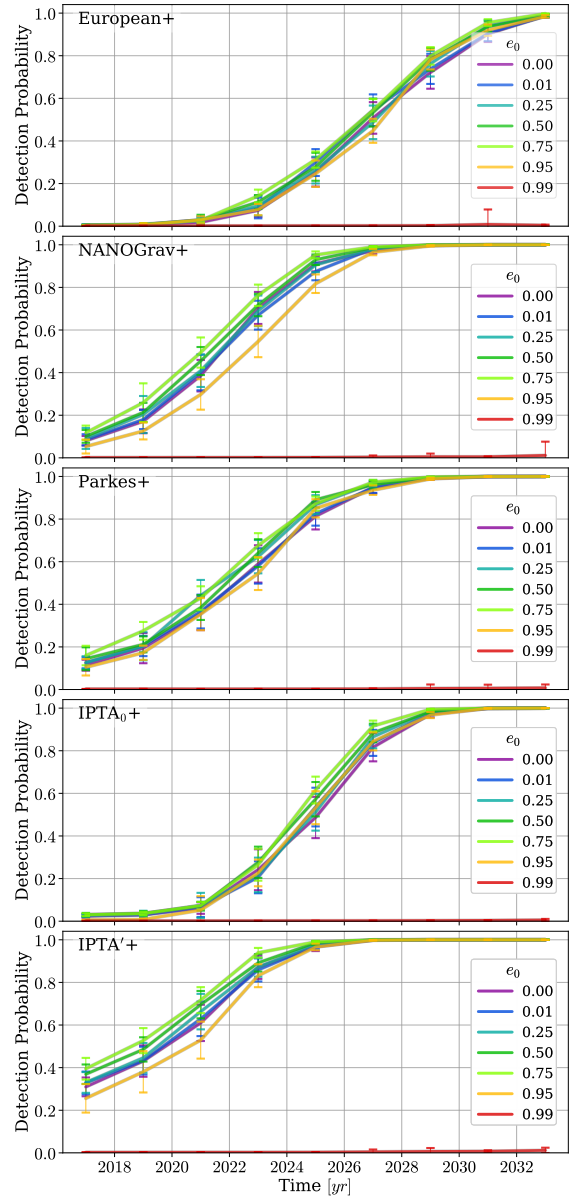}
    \caption{Detection Probability (DP) versus time for eccentric evolution models with different initial eccentricities ($e_0$, colors) and different PTA (rows).  The lines and error bars are averages and standard deviations over 200 MC realizations.  Currently, in 2017, we find detection probabilities are below $20\%$ for all \textit{official} PTA, but reach $95\%$ between about 2025 and 2032.  Higher eccentricities tend to be slightly more detectable, until $e_0 > 0.75$ where the spectral turnover takes a toll on the low-frequency GWB amplitude.}
    \label{fig:dp_ecc-evo}
    \end{figure}

Detection probability versus time are shown for eccentric-evolution models in \figref{fig:dp_ecc-evo}
using expanded PTA configurations.  Averages and standard deviations over 200 MC realizations are plotted.
Ignoring the highest eccentricity case (${e_0 = 0.99}$),
for the moment, the variation between different eccentricities is $\lesssim 20\%$ in DP, and generally
$\simclose 2$~yr at a fixed value.  The different DP growth curves are quite similar between PTA, with
DP tending to be higher for higher eccentricities between ${e_0 = 0.0}$ and ${e_0 = 0.75}$.
Not surprisingly, the ${e_0 = 0.99}$ model is an outlier in DP as in binary evolution.
In this extreme case the GWB is effectively undetectable for all PTA.

The PTA differ in their response to the ${e_0 = 0.95}$ simulations depending on their frequency sensitivities.  Longer observing durations mean PTA are able to detect lower frequencies, and the lowest accessible frequency bins are the most sensitive \citep[see, e.g.,][]{moore2015b}.  Pulsars (and PTA) with the longest observing durations (i.e.~Parkes) are most sensitive to the GWB spectral turnover accentuated in the high eccentricity models.  As observing time increases, however, sensitivity increases at all frequencies, and the addition of short duration (and low noise) expansion pulsars boosts high-frequency sensitivity.  After a sufficient observing time, the high frequency portion of the GWB spectrum, where ${e_0 = 0.95}$ is highest (see \figref{fig:gwb_mc_eccen}), gains enough leverage for its DP to overtake that of lower eccentricities.

In \figref{fig:dp_ecc-evo}, the \iptazero{}+ doesn't perform as well as Parkes+ at early times.  This is due both to I) the differing noise calibrations---the \iptazero{}'s white-noise is pushed significantly higher than Parkes to match the observed upper-limits, as well as, II) the specifications for each individual PTA not quite matching those included in the IPTA public specifications.  Both of these factors motivate our inclusion of the \iprime{} model, which has a higher DP than Parkes, even at early times, as expected.  We are optimistic that future IPTA results and data releases will show even greater improvements than suggested by those of the initial IPTA.

	% ====    FIGURE 9  :  Time to Detection vs. Eccentricity   ====
    \begin{figure}
    \includegraphics[width=\columnwidth]{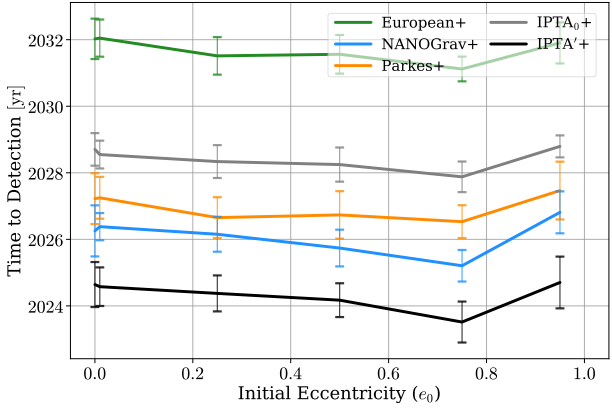}
    \caption{Time to reach a $95\%$ DP versus initial eccentricity ($e_0$) of MBH binaries.  Each (expanded, `+') PTA is shown with the average and standard deviation of 200 MC realizations.  While the official \iptazero{} specification lags behind Parkes and NANOGrav, the more optimistic \iprime{} model reaches $95\%$ DP almost two years earlier.}
    \label{fig:dt_ecc-evo}
    \end{figure}

Figure \ref{fig:dt_ecc-evo} shows times to reach $95\%$ DP versus initial eccentricity for each PTA.  Overall, time to detection tends to improve slightly with increasing eccentricity as it increases the GWB amplitude in the mid-to-upper PTA band.  Differences between eccentricities, however, tend to be comparable or smaller than the variance between MC realizations.  For very high eccentricities, ${e_0 > 0.75}$, the time to detection again increases as the GWB spectral turnover becomes `visible' to PTA, and the signal is diminished.  While the ${e_0 = 0.99}$ models never reach $95\%$ DP in our results, additional simulations of a `rapid' \iprime{}+ model, where expansion pulsars have a cadence of 2 days (see \secref{sec:ipta_models}, ` \iprimerap{} '), are able to reach DP $\sim 50\%$ by 2032.  Even in the highest eccentricity model with ${e_0 = 0.99}$, detection prospects for individual, deterministic sources may not be affected quite as strongly.  We are currently exploring single source predictions from our models, to be presented in a future study.

% DP: CIRCULAR MODELS
% - - - - - - - - - - - -
\subsubsection{Circular Binary Evolution}

	% ====    FIGURE 10  :  Circular Detection Probability vs. Time   ====
	\newcommand{\dpscalenon}{0.95}
    \begin{figure}
    \includegraphics[width=\dpscalenon\columnwidth]{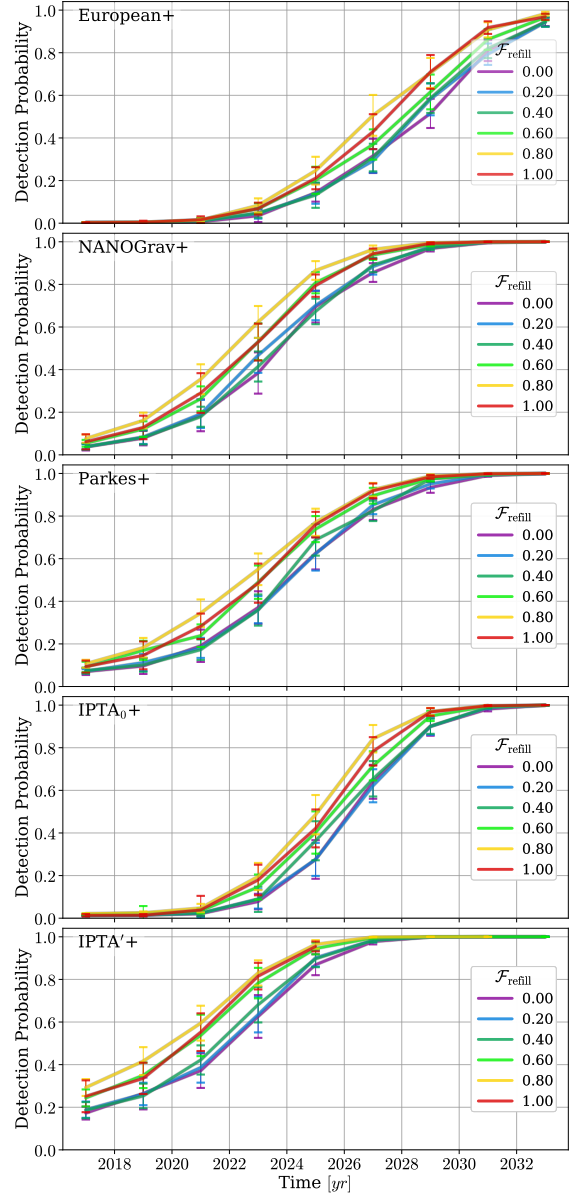}
    \caption{DP versus time for circular evolution models with different LC efficiencies ($\frefill$, colors) and different expanded PTA (rows).  These growth curves behave very similarly to those of the eccentric models, but generally take $\simclose 2$ yr longer to reach the same DP.  There is also a stronger trend across $\frefill$ compared to $e_0$, with ${\frefill=0.0}$ reaching the same DP $\simclose 2$ yr slower than for ${\frefill \sim 0.8}$ -- $1.0$.}
    \label{fig:dp_ecc-non}
    \end{figure}

Detection probability versus time for \textit{circular} evolution models with a variety of loss-cone refilling parameters ($\frefill$) are shown in \figref{fig:dp_ecc-non}.  Recall that a low, `steady-state LC' corresponds to ${\frefill = 0.0}$, and a highly effective, `full LC' to ${\frefill = 1.0}$.  Overall a similar range of durations are required for comparable DP, but the circular models are systematically harder to detect, taking roughly $2$ years longer.  This is not surprising as the eccentric models assume a full LC, and increasing eccentricity tends to further enhance the GWB amplitude in the PTA band.  In general, higher $\frefill$ lead to higher DP after a fixed time.  The circular, full LC, tends to have a slightly lower DP as the attenuation and spectral turnover from stellar scattering take effect, analogous to the highest eccentricities (see, e.g., \figref{fig:gwb_mc_frefill}).  The \iprime{}+ model shows a slight improvement in time to detection between ${\frefill = 0.8}$ and ${\frefill = 1.0}$, suggesting that its high-frequency sensitivity is able to win out.

	% ====    FIGURE 11  :  Time to Detection vs. Loss-Cone Refilling   ====
    \begin{figure}
    \includegraphics[width=\columnwidth]{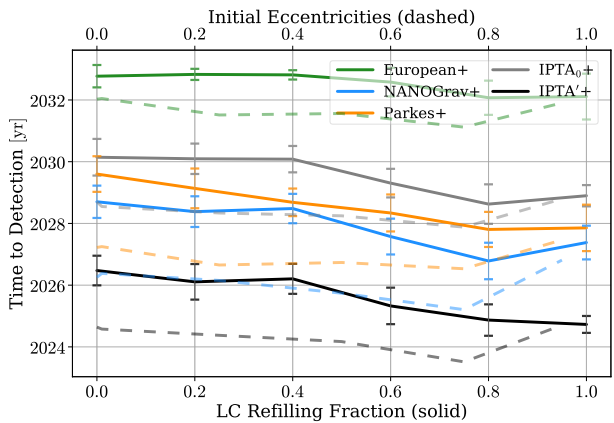}
    \caption{Time to reach a $95\%$ DP versus LC refilling fraction ($\frefill$; solid lines) for circular binary-evolution models.  The times to detection for varying eccentricity models (from \figref{fig:dt_ecc-evo}) are overplotted (dashed lines) for comparison.  Averages and standard deviations are shown.  Varying $\frefill$ has a pronounced effect on detection times, with more effective scattering (larger $\frefill$) models taking less time to observe.  Because the eccentric evolution models assume a full loss-cone, the ${\frefill < 1.0}$ models tend to take longer to be detected.}
    \label{fig:dt_ecc-non}
    \end{figure}

Figure \ref{fig:dt_ecc-non} summarizes the circular-evolution times to detection for each PTA versus $\frefill$ (solid lines).  Overplotted are the eccentric-evolution times to detection (dashed lines, upper x-axis) for comparison.  Based on these trends, a population of MBHB with very high LC scattering efficiency ($\frefill \sim 0.8 - 1.0$) and intermediate eccentricities $0.5 < e_0 < 0.8$ would be the easiest to detect.

% Create a bibliography here, only if just this file is being compiled/built.
\biblio{}

% ===========================================================================================
% = = = = = = = = = = = = = = = = = = = =   OUTRO   = = = = = = = = = = = = = = = = = = = = =
% ===========================================================================================

% Introduction
% ------------

% ==================================================================
% =======================   CONCLUSIONS   =========================

\section{Conclusions}
\label{sec:outro}
This paper has focused on plausible detections of a stochastic Gravitational Wave Background (GWB) using
Pulsar Timing Arrays (PTA).  We have expanded on the Massive Black-Hole (MBH) merger models presented in
\citet{paper1} based on the Illustris simulations.  We have added a model for eccentric binary evolution
assuming `full' Loss-Cone (LC) stellar scattering, in addition to our existing prescriptions for dynamical
friction, stellar scattering with a variety of LC efficiencies ($\frefill$), viscous drag from a
circumbinary disk, and Gravitational Wave (GW) emission.  We have run sets of simulations with a variety of 
LC efficiencies and initial (at the start of stellar scattering) eccentricities.
The MBH binary evolution produced by our models is explored
along with Monte-Carlo (MC) realizations of the resulting stochastic GW Background (GWB) spectra.  Using
parametrized models of currently operational PTA, and their future expansion, we calculate realistic
prospects for detections of the GWB.

The presence of non-zero eccentricity causes two distinct effects to MBH Binary (MBHB) evolution and their
GW spectra.  First, increased eccentricity causes faster GW-hardening and thus more binary coalescence,
but there is additional attenuation of the GWB spectrum and a stronger spectral turnover 
at low frequencies.  Second, while circular binaries emit GW at only
twice the orbital frequency (the $n=2$ harmonic), eccentric binaries also emit at all higher harmonics
(and the $n=1$).  The total GW energy released remains the same, but the overall effect is to move GW
energy from lower to higher frequencies (\figref{fig:gwb_sa_eccen_effects}).

GWB spectra constructed using a Semi-Analytic (SA) calculation (\figref{fig:gwb_sa_eccen}) show
that the amplitude ($\ayr$) of the GWB near the middle of the PTA band ($f\sim 1 \, \pyr$) tends to
increase with increasing eccentricity up to $e_0 = 0.95$.  This is due primarily to the first eccentric
effect: with hardening more effective, the number of binaries coalescing by redshift zero increases.
At lower frequencies, environmental effects---specifically stellar scattering---produce a strong
turnover in the GWB spectrum.  Even moderate eccentricities begin to increase the frequency at which
this turnover occurs because of the second eccentric effect.  Unless the population of binaries
dominating the GWB have very high eccentricities
($e \gtrsim 0.8$), the spectral turnover remains below the PTA sensitive band ($f\lesssim 0.02 \, \pyr$).

The location of the spectral turnover in our models differs from those predictioned by \citet{ravi2014} who
see the turnover at frequencies as high as $f \sim 10^{-1} \, \pyr$.  The location of the
turnover depends on the strength of environmental factors, and thus galactic density profiles.  If
the stellar densities of massive-MBHB host-galaxies are higher than predicted by Illustris, the turnover
could occur at PTA-observable frequencies, regardless of (or exaggerated by) eccentricity distribution.
If the turnover does exist in the current PTA band, it could hurt detection prospects.  At the same time,
observations of such a turnover would be uniquely indicative of environmental interactions, while
observations of the GWB amplitude overall are highly degenerate between cosmological factors
(i.e.~the rate of binary formation) and environmental factors (determining the rate of binary
coalescence).  Currently PTA upper limits of the GWB are still entirely consistent with our results,
and thus are unable to constrain or select between them.

GWB spectra constructed using the MC calculation (\figref{fig:gwb_mc_eccen}), which should resemble real signals, tend to have much steeper strain spectra than $-2/3$ at current PTA frequencies (${f \gtrsim 0.1 \, \pyr}$).  This is because the number of binaries in each frequency bin becomes small, and binary quantization must be taken into account.  MC realizations reveal an interesting corollary to the redistribution of GW energy: with non-zero eccentricity, a larger number of binaries at lower \textit{orbital-frequencies} contribute to the GWB signal at higher \textit{observed-frequency} bins.  This softens the effect of binary quantization and the GWB spectra tend to come closer and closer to a $-2/3$ spectral index with increasing eccentricity (\figref{fig:gwb_mc_sa_rat})---thus producing higher $\ayr$.  For example, the ${e_0 = 0.5}$ \& ${e_0 = 0.95}$ models have $\ayr$, 2~\&~3 times larger than that of the ${e_0 = 0.0}$ model.

To calculate realistic detection statistics, we use parametrized version of each operational PTA: the European (EPTA), NANOGrav, Parkes (PPTA), and International (IPTA; a joint effort of the individual three).  For the IPTA we consider both the public data specifications, \iptazero{}; in addition to a more optimistic, manual combination of the individual groups, \iprime{}.  Overall, our models for NANOGrav, Parkes, and \iptazero{} each behave comparably, reaching $95\%$ detection probability (DP) between 2026 and 2030, and the EPTA following 2 -- 4 years later.  The \iprime{} model noticeably outperforms the others, reaching $95\%$ DP between 2024 and 2026.  High cadence observations of \iprime{} pulsars can further decrease time-to-detection by another $\sim 2$ years.  Moderately high eccentricities ($0.5 \lesssim e_0 \lesssim 0.8$) tend to produce the largest GWB amplitudes in the PTA band.  The eccentricity models used here assume a full loss-cone  (LC; $\frefill \approx 1.0$), and thus circular evolution models which decrease the LC refilling efficiency tend to have lower GWB amplitudes, and longer times to detection by up to $\sim 2$ years.  If galactic-nuclear stellar densities are significantly higher than suggested by Illustris, and the LC is also nearly full, then attenuation of the GWB spectrum could increase times to detection.

Increased eccentricity tends to increase the GWB amplitude, and thus detection prospects.  In the most extreme ${e_0 = 0.99}$ model, however, the signal is so drastically diminished that detections seem unlikely within 20 years in all but the high-cadence IPTA model.  While eccentricities as high as $0.99$ may not be representative of the overall population of binaries, mechanisms which can preferentially drive more massive systems to higher eccentricities \citep[e.g.~counter-rotating stars/gas,][or three-body resonances]{khan2011, amaro-seoane2016} should be further studied.  Varying the eccentricity distribution of binaries has a strong effect on the strain spectral-index of the GWB at high frequencies ($f \gtrsim 1 \, \pyr$).  While PTA are less sensitive at higher frequencies, eventual observations with high signal-to-noise could be used to constrain the underlying binary eccentricity distribution.  The true eccentricity distribution will also affect the prospects for observations of individual, resolvable binaries (`deterministic'/`continuous' sources)---a study of which is currently in progress.

As discussed above, the high frequency portion of the GWB spectrum that PTA will eventually observe is strongly influenced by individual MBHB sources.  In effect, the high frequency portion of the spectrum is no longer a `background'.  It has been shown that this leads to non-Gaussian signal statistics \citep{ravi2012} which are at odds with the assumptions of the detection statistics we use.  Recently, \citet{cornish2016} have found the standard analyses to be robust against small numbers of GW sources.  None the less, if this effect were to systematically decrease GWB detection probabilities, it is likely the effect would be minor because: 1) detection probability is primarily driven at low-frequencies where individual sources are much less important; and 2) the Monte Carlo realization of the spectra we construct should be representative of variations in the GW \textit{background} (neglecting single-sources), and thus our DP and time-to-detection error bars should still be representative.

For a given PTA configuration, the differences in times-to-detection for varying GWB model parameters are at most a few years.  This result is promising as it suggests that, despite uncertainties in the underlying physical processes of binary mergers, the expectation of GWB detections in the near future remains robust.  At the same time it begs the question, `will PTA be able to discern between different models in their observations?'  Based only on the \textit{overall GWB amplitude} (or equivalently the time-to-detection), only a mixed measurement of the overall merger process and the typical MBH binary mass distribution will be constrained.  The different hardening models are largely degenerate in the overall GWB amplitude they predict, especially when taking into account uncertainties in cosmological factors---most notably the true, unbiased distribution of MBH masses \citep[e.g.][]{shen2008, shankar2016}---which is outside of the scope of this study\footnote{
Examining the effects of varying Illustris MBH evolution and masses is being examined.  The MBH population from Illustris---especially at the high-mass end which most strongly effects the GWB---is tightly constrained by the M-$\sigma$ relation and AGN luminosity function which are both accurately reproduced \citep{sijacki2015}.}.

Once PTA have detected the GWB, and signal-to-noise continues to grow, the \textit{shape} of the GWB will be measured which encodes very detailed information about the merger process and typical MBH environments \citep[e.g.][]{taylor2017, chen2017}.  The \textit{strength} of the (low-frequency) spectral turnover is determined by the MBHB coupling to their local stellar environments, and its \textit{location} is additionally effected by the eccentricity distribution of binaries.  The (high-frequency) spectral-index, however, measures the number of sources contributing to the GWB, and thus the underlying eccentricity distribution.  In the ideal, high signal-to-noise regime, the spectral index will determine typical binary eccentricities which can then be disentangled from the stellar coupling, measured from the spectral turnover.  With eccentricity and the loss-cone constrained, the typical masses of merging MBHB can then be inferred from the overall GWB amplitude.

Low frequency sensitivity, established by long observing baselines, tends to drive increases in detection probability.  Still, we find that including short cadence observations to maintain or improve high frequency sensitivity can make a noticeable difference in detection prospects, especially for the most extreme hardening and eccentricity models (in which the GWB spectra turn over at low frequencies).  Regardless of cadence, we find that the continued addition of pulsars monitored by PTA is \textit{essential} for a detection to be made within the next 20 years.  Across a wide range of specific configurations, and even with pessimistic model parameters, if PTA continue to expand as they are, GWB detections are highly likely within about 10 years.

% Create a bibliography here, only if just this file is being compiled/built.
\biblio{}

% Acknowledgments
% ---------------
\section*{Acknowledgments}
We are grateful to Pablo Rosado who was extremely helpful in clarifying details of the PTA detection statistics, and to the referee for thorough and wholly constructive feedback which significantly improved this paper.

This research made use of \astropy, a community-developed core Python package for Astronomy \citep{astropy2013},
in addition to \scipy~\citep{scipy}, \ipython~\citep{ipython}, \numpy~\citep{numpy2011}.  All figures were
generated using \matplotlib~\citep{matplotlib2007}.

% Bibliography
% ------------

\let\oldUrl\url
\renewcommand{\url}[1]{\href{#1}{Link}}

\quad{}
\bibliographystyle{mnras}
\bibliography{refs}

% Appendices
% ----------

\appendix

% ==================================================
% =============   ADDITIONAL EQUATIONS   ============
\section{Additional Equations}
\label{sec:gwb_calc}

The GW frequency distribution function can be expressed as \citep[][Eq.~20]{peters1963},
	\begin{equation}
	\label{eq:gw_freq_dist_func}
	\begin{gathered}
	g(n,e) \equiv \frac{n^4}{32} \left(\left[G_1\vphantom{e^2}\right]^2 + 
	\left[1-e^2\right] \left[G_2\vphantom{e^2}\right]^2 + \frac{4}{3n^2}\left[J_n(ne)\vphantom{e^2}\right]^2 \right), \\
	G_1(n,e) \equiv J_{n-2}(ne) - 2eJ_{n-1}(ne) + \frac{2}{n} J_n (ne) + 2e J_{n+1} (ne) - J_{n+2} (ne) \vphantom{\frac{2}{n}}, \\
	G_2(n,e) \equiv J_{n-2}(ne) - 2eJ_n(ne) + J_{n+2}(ne)\vphantom{\frac{2}{n}}.
	\end{gathered}	
	\end{equation}
Here $J_n(x)$ is the n'th Bessel Function of the first kind.  The sum of all harmonics,
${\sum_{n=1}^\infty g(n,e) = F(e)}$, where $F(e)$ is defined in \refeq{eq:fe}.

The observed, characteristic strain from a set of individual sources is
\citep[e.g.][Eq.~8]{rosado2015},
	\begin{equation}
	\label{eq:strain_from_sources}
	h_\trt{c}^2 = \sum_i h_\trt{s,i}^2 \, \frac{f_i}{\Delta f} \approx \sum_i h_\trt{s,i}^2 \, f_i \, T,
	\end{equation}
where the second equality assumes that frequency bins are determined by the resolution corresponding to
a total observational duration $T$.

The cosmological evolution function is \citep[Eq.~14]{hogg1999},
	\begin{equation}
	\label{eq:cosmo_func}
    E(z)\equiv\sqrt{\Omega_{\rm M}\,(1+z)^3+\Omega_k\,(1+z)^2+\Omega_{\Lambda}},
	\end{equation}
for $z$ the redshift, and $\Omega_{\rm M}$, $\Omega_k$ \& $\Omega_{\Lambda}$ the density parameters
for matter, curvature and dark-energy.

% =============   DETECTION FORMALISM   ============
\subsection{Detection Formalism}
\label{sec:ap_det}
In what follows, the GWB signal is characterized by a Spectral Energy Density (SED),
	\begin{equation}
	S_h = \frac{h_c^2}{12 \pi^2 f^3},
	\end{equation}
and the prediction/model SED is denoted as $S_{h0}$.  In all of our calculations, we use
a purely power-law GWB spectrum for $S_{h0}$ with an amplitude of $\ayr = 10^{-16}$.
Each pulsar $i$ is characterized by a noise SED $P_i$ (\refeq{eq:total_noise_power}).

PTA detection statistics typically rely on cross-correlations between signals using an
`overlap reduction function' \citep[the][curve]{hellings1983},
	\begin{equation}
	\Gamma_{ij} = \frac{3}{2} \gamma_{ij} \, \ln \left(\gamma_{ij}\right) - \frac{1}{4}\gamma_{ij} + \frac{1}{2} + \frac{1}{2}\delta_{ij}
	\end{equation}
where,
	\begin{equation}
	\gamma_{ij} = \frac{1}{2}\left[1 - \cos(\theta_{ij})\right],
	\end{equation}
for an angle between pulsars $i$ and $j$, $\theta_{ij}$.

We employ the `B-Statistic' from \citet{rosado2015}, constructed by maximizing the statistic's
SNR---defined as the expectation value of the statistic in the presence of a signal,
	\begin{equation}
	\mu_{B1} = 2 \pulsarsum \frac{\Gamma_{ij}^2 \, S_h \, S_{h0}}{\lr{P_i + S_{h0}} \lr{P_j + S_{h0}} + \Gamma_{ij}^2 \, S_{h0}^2},
	\end{equation}
divided by the variance of the statistic also in the presence of a signal,
	\begin{equation}
	\sigma_{B1}^2 = 2 \pulsarsum \frac{\Gamma_{ij}^2 \, S_{h0}^2 \left[ \lr{P_i + S_h} \lr{P_j + S_h} + \Gamma_{ij}^2 \, S_h^2\right]}{\left[{\lr{P_i + S_{h0}} \lr{P_j + S_{h0}} + \Gamma_{ij}^2 \, S_{h0}^2}\right]^2},
	\end{equation}
i.e.~$\SNR_B \equiv \mu_{B1} / \sigma_{B1}$, as apposed to the variance in the \textit{absence} of a signal,
	\begin{equation}
	\sigma_{B0}^2 = 2 \pulsarsum \frac{\Gamma_{ij}^2 \, S_{h0}^2 P_i \, P_j}{\left[{\lr{P_i + S_{h0}} \lr{P_j + S_{h0}} + \Gamma_{ij}^2 \, S_{h0}^2}\right]^2}.
	\end{equation}
The SNR can then be expressed as,
	\begin{equation}
	\label{eq:snr_b}
	\SNR^2 = \SNR_B^2 = 2 \pulsarsum \frac{\Gamma_{ij}^2 \, S_h^2}{P_i P_j + S_h\lr{P_i + P_j} + S_h^2\lr{1 + \Gamma_{ij}^2}},
	\end{equation}
which is only meaningful compared to the threshold-SNR for a particular false-alarm probability ($\alpha_0$)
and DP-threshold ($\gamma_0$),
	\begin{equation}
	\label{eq:snr_thresh_b}
	\SNR_B^T = \sqrt{2} \left[ \frac{\sigma_0}{\sigma_1} \erfcinv{2\alpha_0} - \erfcinv{2\gamma_0} \right].
	\end{equation}
The SNR can be circumvented altogether by considering the measured DP,
	\begin{equation}
	\label{eq:det_prob_b}
	\gamma_B = \frac{1}{2} \erfc{ \frac{\sqrt{2} \, \sigma_0 \erfcinv{2\alpha_0} - \mu_1}{\sqrt{2} \, \sigma_1} },
	\end{equation}
which is the primary metric we use throughout our analysis of PTA detections.

% ===========================================================
% =============   BINARY HOST-GALAXY DENSITIES   ============
%\newpage
\section{Host-Galaxy Densities in Illustris}
\label{sec:ap_gal_dens}
The point at which two MBH come within a smoothing length of one another is identified in Illustris,
and density profiles are calculated for the host galaxy at that time.  The profiles are
used to calculate the environmental hardening rates which then determine the GWB spectra.  In
particular, the stellar densities strongly affect the location of the spectral turnover through
stellar scattering.  Because the location of the spectral turnover is especially important for future
detections of the GWB, we provide some additional details on the stellar environments here.

\begin{figure}
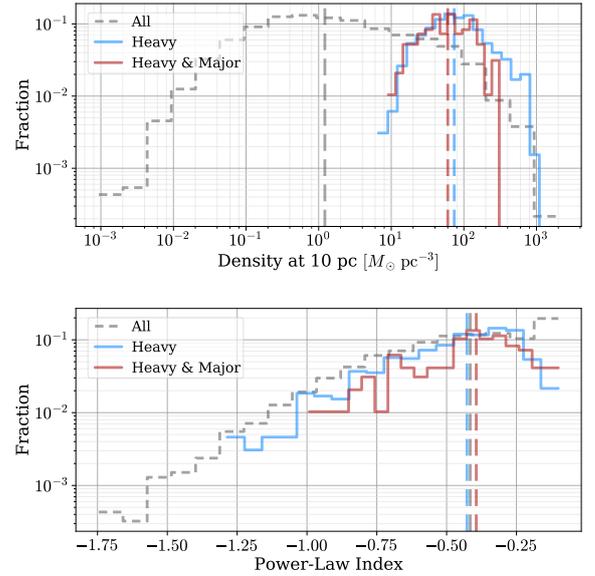

	\subfloat{\includegraphics[width=0.95\columnwidth]{../figs/{fit_dens_at_10pc}}} \vspace{-10pt}

	\subfloat{\includegraphics[width=0.95\columnwidth]{../figs/{fit_pwl_index}}}
    \caption{Upper-panel: distribution of stellar densities at $10 \, \pc$ for Illustris galaxies
    hosting MBH binaries.  For galaxies in which $10 \, \pc$ is unresolved, the density is calculated from
    power-law fits to the inner-most (resolved) regions.  The dashed grey lines show the entire
    population of binary host galaxies,
    while the blue lines show hosts of \heavydef{} binaries, and red the \heavy{} and \majordef{}
    binaries.  Each population is plotted fractionally, but note that \heavy{} binaries constitute
    $\closesim 7\%$ and \heavymajor{} $\closesim 1\%$ of all binaries respectively.  Vertical lines indicate the
    median value of each subset.}
    \label{fig:gal_dens}
	\end{figure}

To calculate density profiles, we average the density of each particle type (star, dark matter, and gas)
in radial bins.  Because Illustris is only able to resolve down to 10s--100s of parsec scales, we extrapolate to
smaller radii with power-law fits to the eight inner-most bins\footnote{Restricted to those which contain
at least four particles each.}.  
\figref{fig:gal_dens} shows the distribution of stellar densities at 10 pc (interpolated or extrapolated as
needed) in the upper-panel\footnote{We choose 10 pc as it is near typical spheres of influence ($\rinfl$) \&
hardening radii ($\rh$) for our systems, observational resolution-limits for nearby galaxies, and usually just
beneath Illustris resolution-limits.}, and power-law indices in the lower-panel.  The overall population
of binaries are shown in grey (dashed), in addition to the \heavydef{}
subset in blue, and \heavy{} \& \majordef{} mass-ratio subset in red.  There is a roughly 100 times increase in
typical stellar densities between \heavy{} systems and overall host-galaxies, but no noticeable change when
further selecting by mass-ratio.  While the \heavy{} subset constitutes less than $10\%$ of systems,
they contribute $\closesim 90\%$ of the GWB amplitude \citepalias[see,][]{paper1}.

The median power-law index for the inner stellar density profiles is $\simclose -0.4$.  For comparison, at
small radii an \citet{hernquist1990} profile corresponds to $-1$, and $-1.5$ produces a surface-density
distribution that resembles a \citet{devaucouleurs1948} profile \citep{dehnen1993}.  At the
same time, many massive galaxies (comparable to our host galaxies) have flattened `cores` in their stellar
density profiles \citep[e.g.][]{faber1997, lauer2007} and it has long been proposed that these cores could
be explained by dynamical scouring from MBH binaries \citep[e.g.][]{quinlan1997, volonteri2003}.  Both 
computationally and observationally, inner density profiles in the `hard' binary regime (typically
${r \lesssim 10 \, \pc}$) are very difficult to resolve.  It is thus unclear how accurate these profiles are.
While they may be realistic models, some of the flattening in the inner regions may be due in part to numerical 
effects (e.g.~gravitational softening in the force calculations) or the known, over-inflated
radii of some galaxies in Illustris \citep{snyder2015, paper1}.

% ====================================================
% =============   C: ADDITIONAL FIGURES   ============
\section{Additional Figures}

	% ====    FIGURE C1 : Det Prob Models - PTA Grid    ====
    \begin{figure*}
    \includegraphics[width=0.96\textwidth]{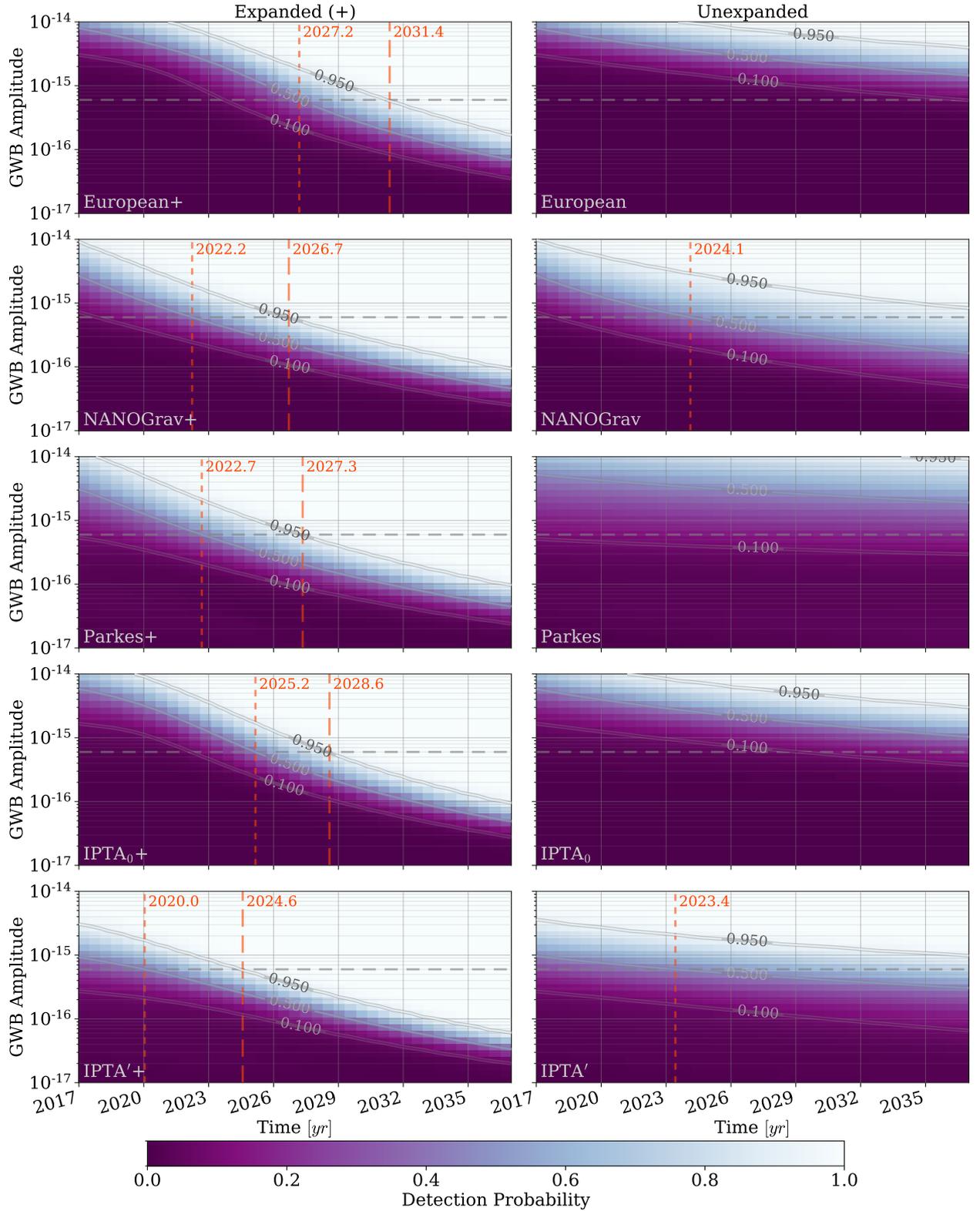} 
    \caption{Detection Probability (DP) for purely power-law GWB spectra exploring different intrinsic GWB
    amplitudes over different observation durations for each PTA.  The left column shows the expanded
    (`+') configurations where new pulsars are added each year, while the right column shows the static
    configurations with only the current number of pulsars.  The IPTA$_0$ is the official specification for the
    International pulsar timing array, while the \iprime{} is a more optimistic, manual combination
    of the specifications for each of the three individual PTA (see \secref{sec:meth_pta}).
    The horizontal, dashed grey lines show the GWB
    amplitude from our fiducial model: $\ayr = 0.6\E{-15}$, and the vertical, dashed orange lines show the
    time at which each configuration reaches $\mathrm{DP} = 50\%$ (short-dashes) and $\rm{DP} = 95\%$
    (long-dashes) for the fiducial amplitude.  For a power-law spectrum at the fiducial amplitude, we expect
    \iprime{}+ to reach $50\%$ \& $95\%$ DP in about 3 \& 8 years ($\sim 2020$ \& $\sim2025$) respectively.  Without expansion, the \iprime{} reached $50\%$ DP in about 6 years ($\sim2023$), and does not reach $95\%$ DP within 20 years.}
    \label{fig:dp_ppl}
    \end{figure*}

	% ====    FIGURE C2 : Eccentricity GWB Ratio - MC/SA    ====
    \begin{figure}
    \centering
    \includegraphics[width=\columnwidth]{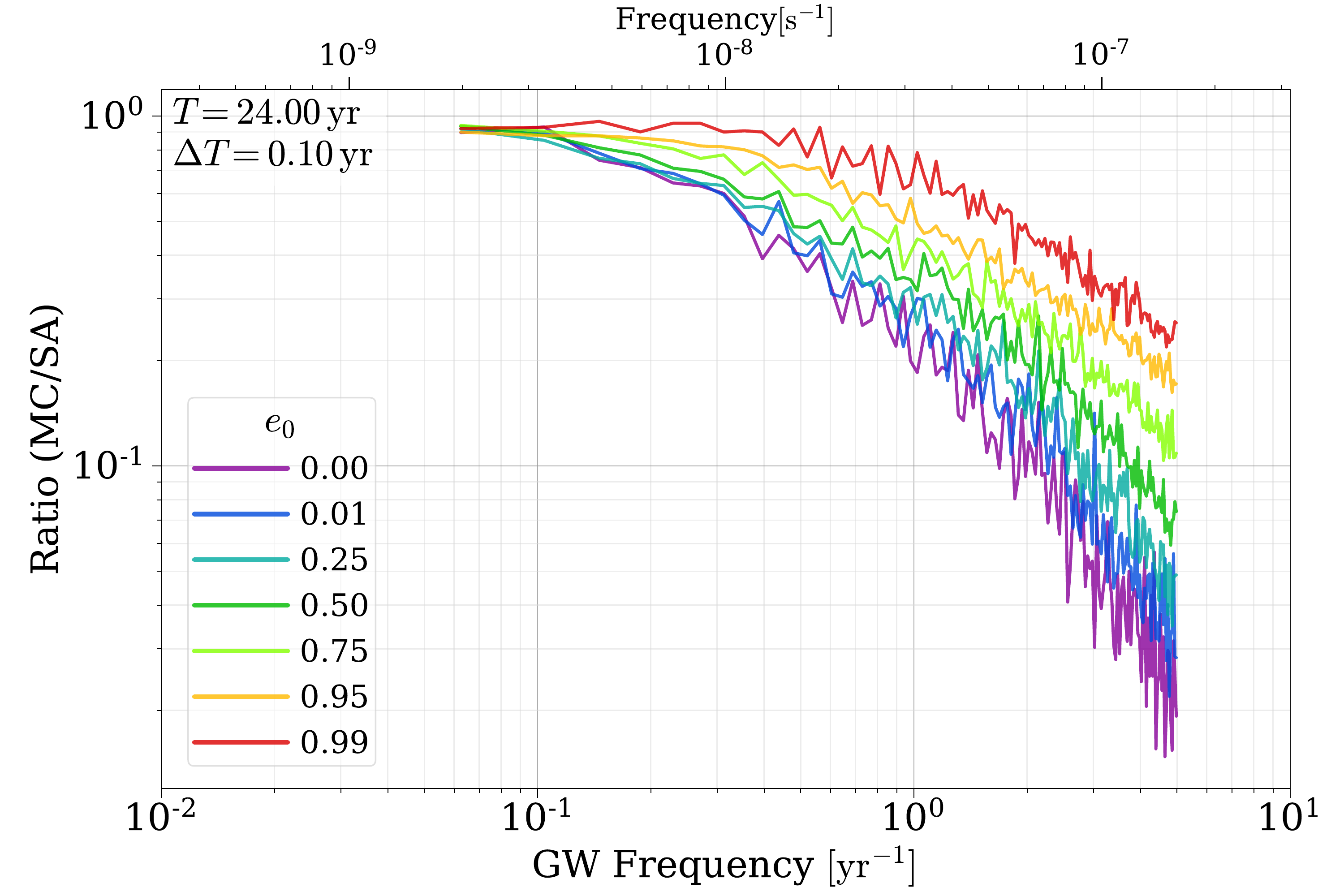}
    \caption{Ratio of the (median) MC-calculated GWB to that of the SA calculation.  Binaries at higher
    eccentricities contribute more GW energy to higher-harmonics above their orbital frequency.  This
    causes the number of sources contributing at higher-frequencies to increase with increasing
    eccentricity, decreasing the effects of MBHB quantization.}
    \label{fig:gwb_mc_sa_rat}
    \end{figure}

	% ====    FIGURE C3 : Circular GWB - SA & MC for Refilling Ratios    ====
    \begin{figure}
    \centering
    \includegraphics[width=\columnwidth]{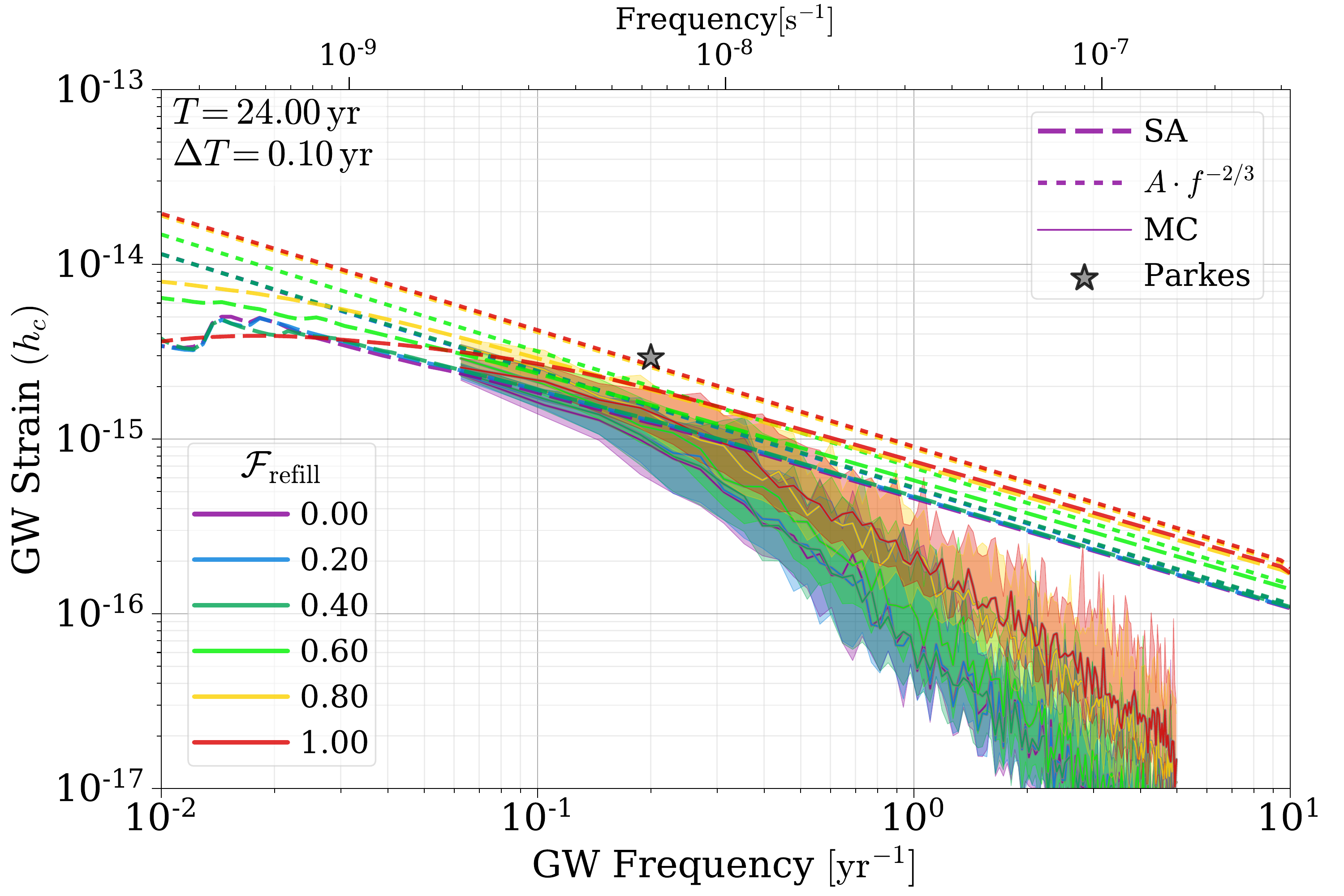}
    \caption{Monte-Carlo (MC) calculated GWB for a variety of different LC refilling fractions ($\frefill$),
    with median lines (solid) and one-sigma contours shown.  Both the full (dashed) and power-law
    (dotted) SA models are also plotted.  More efficient LC refilling means more binaries coalesce, causing the
    GWB amplitude to increase.  An always full LC causes increased attenuation at lower frequencies: apparent
    at $f\lesssim0.1 \, \pyr$.  The steepening of the spectral index at higher frequencies due to finite-number
    effects is also apparent, but for $\frefill >= 0.8$, the effect is somewhat remediated.}
    \label{fig:gwb_mc_frefill}
    \end{figure}

	% ====    FIGURE C4 : Circular GWB Ratio - MC/SA    ====
    \begin{figure}
    \centering
    \includegraphics[width=\columnwidth]{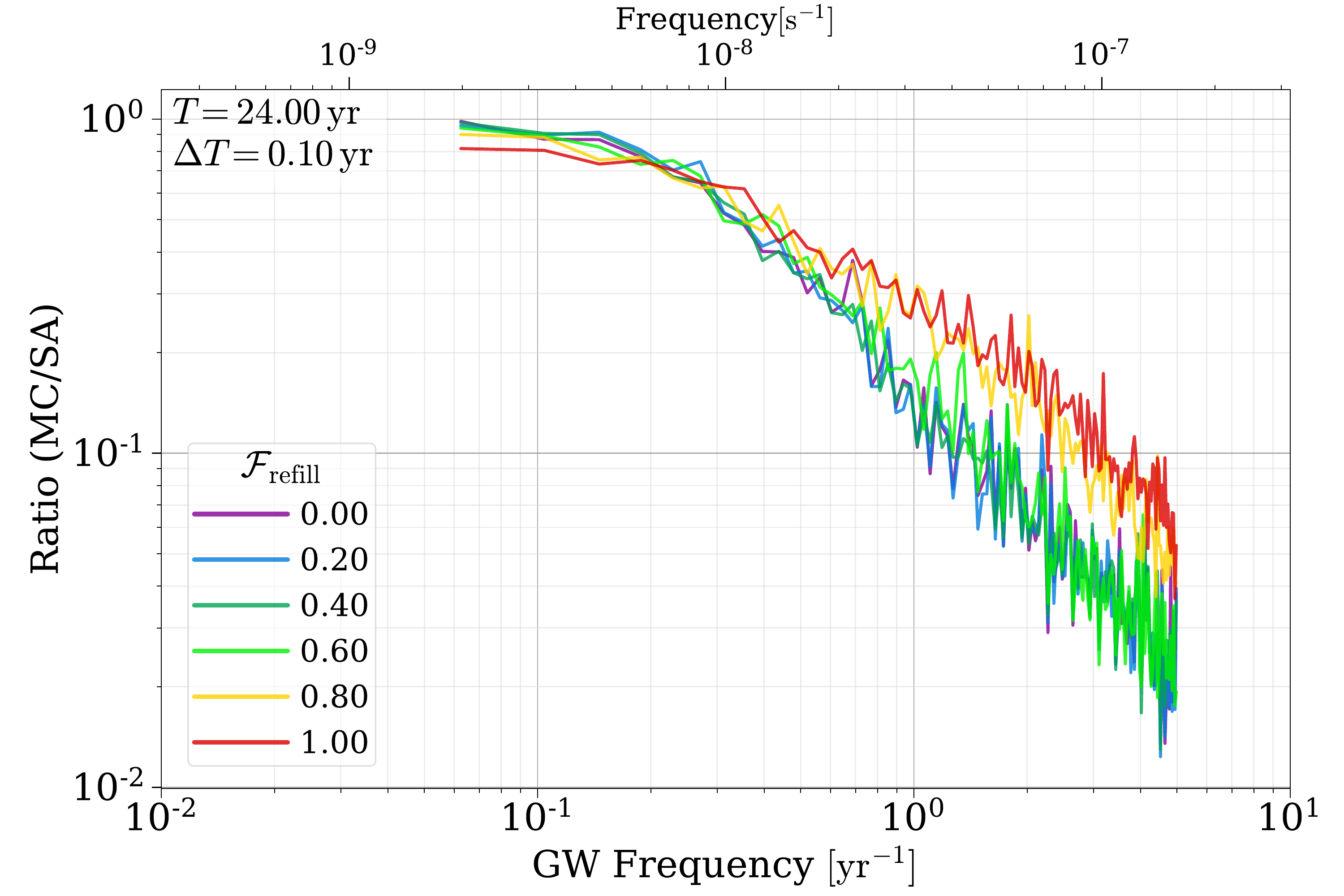}
    \caption{Ratio of the (median) MC-calculated GWB to that of the SA calculation, for zero-eccentricity
    and a variety of LC refilling fractions ($\frefill$).  Finite-number effects, from few binaries in each
    bin, cause the strong deviation between Semi-Analytic (SA) and MC calculations at higher frequencies.
    This effect is somewhat alleviated by effective LC refilling ($\frefill >= 0.8$) where
    the total number of coalescing binaries is increased.}
    \label{fig:gwb_mc_sa_rat_frefill}
    \end{figure}

% =============================================================================
% =============   D: INTERNATIONAL PTA, AND MODEL SENSITIVITIES    ============
%\newpage
\section{International PTA Models and Time-to-Detection Sensitivities}
\label{sec:ipta_models}

	% ====    FIGURE D1 : Det Prob Models - International PTA Grid    ====
    \begin{figure*}
    \includegraphics[width=0.96\textwidth]{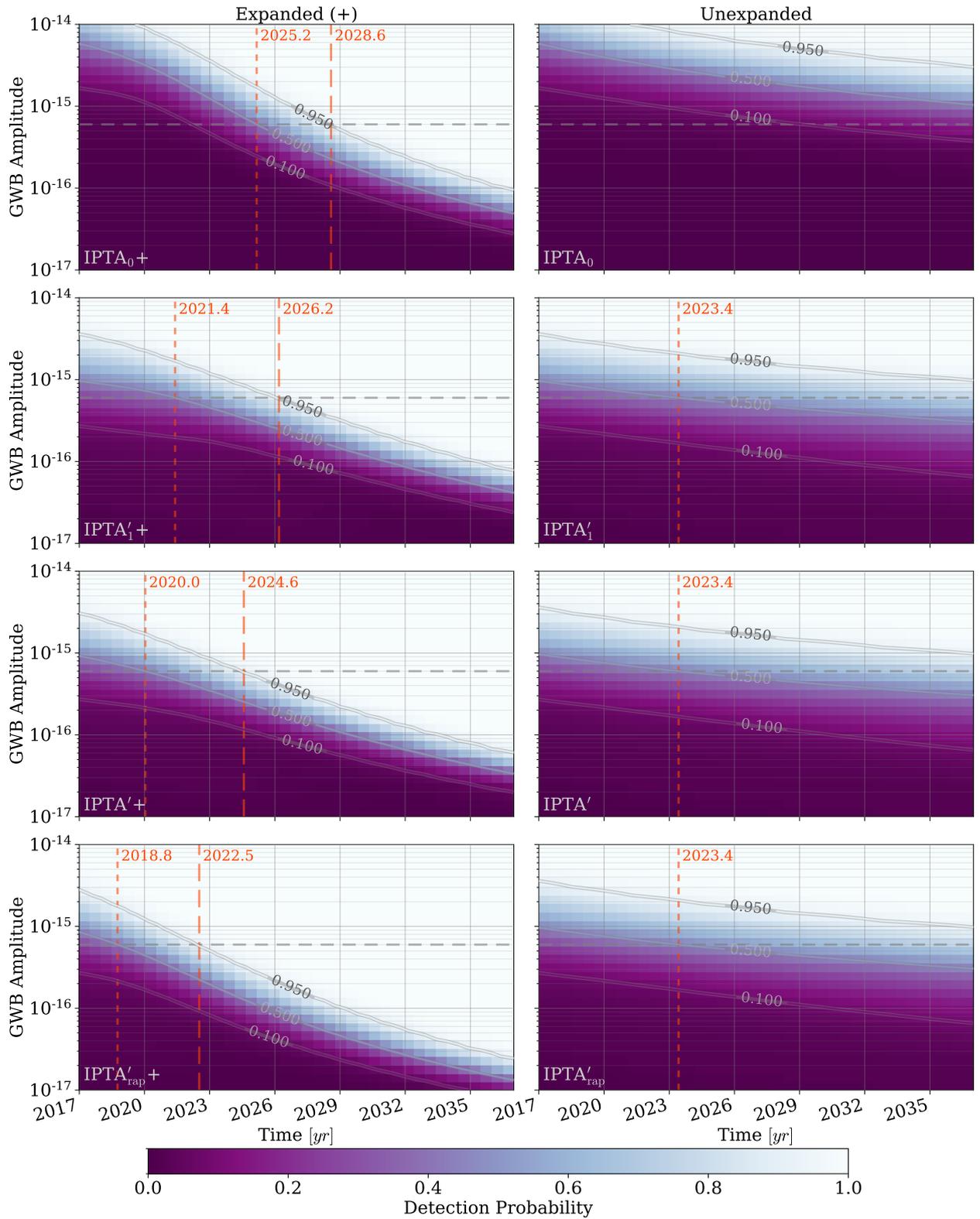} 
    \caption{Detection probability for purely power-law GWB spectra of varying amplitudes versus observing time.  Shown are the four different International PTA configurations discussed above, with differences in the pulsar characteristics (\iptazero{} versus the other models) and different expansion specifications for adding pulsars (\iprimeone{}+, \iprime{}+, and \iprimerap{}+).}
    \label{fig:dp_ppl_ipta}
    \end{figure*}

In this section we discuss different aspects of models for the International PTA.  The \iptazero{} model is based on the official, public IPTA data release \citep{verbiest2016}.  Throughout this paper we have also focused on the \iprime{} model (discussed in \secref{sec:meth_pta}), which is a manual combination of the public data sets from each of the three individual PTA.  \tabref{tab:pta} shows a summary of the differences between these primary PTA models.  The time at which PTA will make detections depends sensitively on how they expand: how rapidly they add new pulsars to the arrays, and what the timing parameters of those pulsars are.  We use `expanded' PTA models (denoted with a `+') to account for this growth.  The \iprime{}+ model gradually increases the rate of expansion from 2011 to 2015, to account for the staggered ends of the individual PTA data sets.  Initially the \iprime{}+ expands by 2 pulsars per year (after 2011), and finally by 6 per year (after 2015).  Here, we also introduce a \iprimeone{}+ model which does not expand at all until after 2015, at which point it adds 6 pulsars per year.  Finally, we also show a model which uses the same expansion schedule as \iprime{}+, but in which the pulsars added have a rapid cadence of 2 days, called \iprimerap{}+.

Figure \ref{fig:dp_ppl_ipta} shows detection probability for different purely power-law GWB amplitudes for the different IPTA models.  The \iptazero{} and \iprimeone{} differ in the overall pulsar parameters (noise, cadence, etc).  In the expanded cases, the differences in \iptazero{}+ and \iprimeone{}+ models lead to differences of 4 \& 2.5 years to reach $50\%$ \& $95\%$ DP respectively for our fiducial amplitude of ${\ayr = 0.6\E{-15}}$.  In the unexpanded models the difference is even more pronounced where by 2037 the \iptazero{} hardly reaches $50\%$ DP for an amplitude of ${\ayr = 10^{-15}}$, while \iprimeone{} reaches $50\%$ DP for ${\ayr = 0.6\E{-15}}$ in $\sim 2023$.

The \iprimeone{}+, \iprime{}+, and \iprimerap{}+ models differ in only their expansion specifications so their detection probabilities in the unexpanded configurations are identical.  The expanded versions however differ notably.  \iprimeone{}+ vs.~\iprime{}+ (expanding after 2015 vs.~gradually increasing expansion starting in 2011) differ in time to detection by $\sim 1.5$ yr for both $50\%$ and $95\%$ DP (again at $\ayr = 0.6\E{-15}$).  Going from the \iprime{}+ model to the \iprimerap{}+ model (decreasing the observing cadence of added pulsars from every $\sim14$ days to every 2 days) further decreases the time to detection by 1 \& 2 years for $50\%$ and $95\%$ DP.

\begin{figure}
    \centering
    \includegraphics[width=\columnwidth]{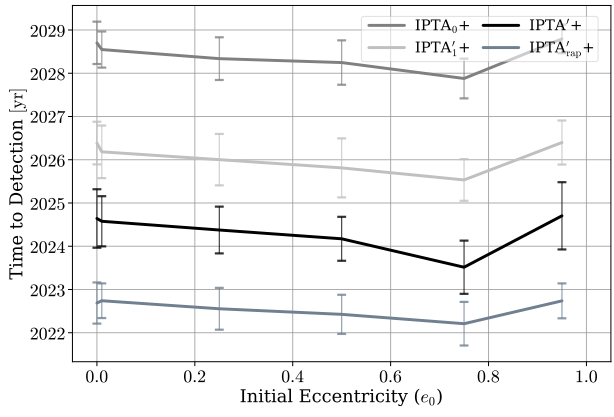}
    \caption{Time to detection versus initial eccentricity for the same four International PTA configurations discussed above.  Differences in pulsar characteristics (most notably noise properties) and expansion prescriptions yield a 6 year range in times to detection.}
    \label{fig:dt_ecc-evo_ipta}
    \end{figure}

Figure \ref{fig:dt_ecc-evo_ipta} shows time to detection (at $95\%$ DP) versus initial eccentricity for the full GWB calculation.  Overall, differences between IPTA models lead to a 6 year range of possible times-to-detection for the same GWB spectra.  This highlights 1) \textbf{the importance of the red-noise characterization of pulsars}, which often disagree significantly between different PTA but for the same pulsar; 2) that the expansion prescriptions we are using are \textit{ad hoc}, and \textbf{updates from the individual PTA and especially the IPTA on their current data sets are very important} moving ahead; and 3) that \textbf{higher cadence observations (i.e.~more telescope time) will make a noticeable improvement in time to detection} (and likely how quickly SNR will grow after detection) even without consider the benefits to noise characterization.

	% ====    FIGURE D2 : Time to Detection - International PTA     ====

% Create a bibliography here, only if just this file is being compiled/built.
\biblio{}

\label{lastpage}

\end{document}